\documentstyle[chicagob,12pt]{article}

\newtheorem{THEOREM}{Theorem}[section]
\newenvironment{theorem}{\begin{THEOREM} \hspace{-.85em} {\bf :} }%
                        {\end{THEOREM}}
\newtheorem{LEMMA}[THEOREM]{Lemma}
\newenvironment{lemma}{\begin{LEMMA} \hspace{-.85em} {\bf :} }%
                      {\end{LEMMA}}
\newtheorem{COROLLARY}[THEOREM]{Corollary}
\newenvironment{corollary}{\begin{COROLLARY} \hspace{-.85em} {\bf :} }%
                          {\end{COROLLARY}}
\newtheorem{PROPOSITION}[THEOREM]{Proposition}
\newenvironment{proposition}{\begin{PROPOSITION} \hspace{-.85em} {\bf :} }%
                            {\end{PROPOSITION}}
\newtheorem{DEFINITION}[THEOREM]{Definition}
\newenvironment{definition}{\begin{DEFINITION} \hspace{-.85em} {\bf :} \rm}%
                            {\end{DEFINITION}}
\newtheorem{CLAIM}[THEOREM]{Claim}
\newenvironment{claim}{\begin{CLAIM} \hspace{-.85em} {\bf :} \rm}%
                            {\end{CLAIM}}
\newtheorem{EXAMPLE}[THEOREM]{Example}
\newenvironment{example}{\begin{EXAMPLE} \hspace{-.85em} {\bf :} \rm}%
                            {\end{EXAMPLE}}
\newtheorem{REMARK}[THEOREM]{Remark}
\newenvironment{remark}{\begin{REMARK} \hspace{-.85em} {\bf :} \rm}%
                            {\end{REMARK}}

\newcommand{\thm}{\begin{theorem}}
\newcommand{\lem}{\begin{lemma}}
\newcommand{\pro}{\begin{proposition}}
\newcommand{\dfn}{\begin{definition}}
\newcommand{\rem}{\begin{remark}}
\newcommand{\xam}{\begin{example}}
\newcommand{\cor}{\begin{corollary}}
\newcommand{\prf}{\noindent{\bf Proof:} }
\newcommand{\ethm}{\end{theorem}}
\newcommand{\elem}{\end{lemma}}
\newcommand{\epro}{\end{proposition}}
\newcommand{\edfn}{\bbox\end{definition}}
\newcommand{\erem}{\bbox\end{remark}}
\newcommand{\exam}{\bbox\end{example}}
\newcommand{\ecor}{\end{corollary}}
\newcommand{\eprf}{\bbox\vspace{0.1in}}
\newcommand{\beqn}{\begin{equation}}
\newcommand{\eeqn}{\end{equation}}

\newcommand{\bbox}{\vrule height7pt width4pt depth1pt}

\newcommand{\clm}{\begin{claim}}
\newcommand{\eclm}{\end{claim}}

\newcommand{\bor}{\bigvee}

\newcommand{\union}{\cup}
\newcommand{\inter}{\cap}

\newcommand{\IR}{\mbox{$I\!\!R$}}

\newcommand{\IN}{\mbox{$I\!\!N$}}

\renewcommand{\phi}{\varphi}

\newcommand{\infinity}{\infty}

\newcommand{\A}{{\cal A}}

\newcommand{\C}{{\cal C}}

\newcommand{\F}{{\cal F}}
\newcommand{\G}{{\cal G}}

\newcommand{\R}{{\cal R}}
\newcommand{\T}{{\cal T}}
\newcommand{\U}{{\cal U}}
\newcommand{\V}{{\cal V}}
\newcommand{\W}{{\cal W}}

\newcommand{\<}{\langle}
\renewcommand{\>}{\rangle}

\newcommand{\ie}{i.e.,~}

\newcommand{\respc}{resp.,\ }

\newcommand{\ol}{\setlength{\itemsep}{0pt}\begin{enumerate}}
\newcommand{\eol}{\end{enumerate}\setlength{\itemsep}{-\parsep}}
\newcommand{\ul}{\setlength{\itemsep}{0pt}\begin{itemize}}
\newcommand{\dl}{\setlength{\itemsep}{0pt}\begin{description}}
\newcommand{\edl}{\end{description}\setlength{\itemsep}{-\parsep}}
\newcommand{\eul}{\end{itemize}\setlength{\itemsep}{-\parsep}}

\newcommand{\commentout}[1]{}

\newcommand{\bi}{\begin{itemize}}
\newcommand{\ei}{\end{itemize}}
\newcommand{\be}{\begin{enumerate}}
\newcommand{\ee}{\end{enumerate}}

\newenvironment{oldthm}[1]{\par\noindent{\bf Theorem #1:} \em \noindent}{\par}
\newenvironment{oldlem}[1]{\par\noindent{\bf Lemma #1:} \em \noindent}{\par}
\newenvironment{oldcor}[1]{\par\noindent{\bf Corollary #1:} \em \noindent}{\par}
\newenvironment{oldpro}[1]{\par\noindent{\bf Proposition #1:} \em \noindent}{\par}
\newcommand{\othm}[1]{\begin{oldthm}{\ref{#1}}}
\newcommand{\eothm}{\end{oldthm} \medskip}
\newcommand{\olem}[1]{\begin{oldlem}{\ref{#1}}}
\newcommand{\eolem}{\end{oldlem} \medskip}
\newcommand{\ocor}[1]{\begin{oldcor}{\ref{#1}}}
\newcommand{\eocor}{\end{oldcor} \medskip}
\newcommand{\opro}[1]{\begin{oldpro}{\ref{#1}}}
\newcommand{\eopro}{\end{oldpro} \medskip}

\newcommand{\tends}{\rightarrow}
\newcommand{\tendsto}{\tends}

\newcommand{\bxor}[1]{\dot{\bor}}

\newcommand{\aeq}{\approx} 

\newcommand{\naeq}{\not\approx}

\newcommand{\vecmu}{\vec{\mu}}
\newcommand{\Supp}{\mbox{\it Supp}}
\newcommand{\FCP}{F_{S \rightarrow P}}
\newcommand{\FOP}{F_{O \rightarrow C}}
\newcommand{\FDC}{F_{D \rightarrow C}}
\newcommand{\FLN}{F_{L \rightarrow N}}
\newcommand{\FSN}{F_{S \rightarrow N}}
\newcommand{\FNP}{F_{N \rightarrow P}}
\newcommand{\LPS}{\mbox{\em LPS\/}}
\newcommand{\SLPS}{\mbox{\em SLPS\/}}
\newcommand{\NPS}{\mbox{\em NPS\/}}
\newcommand{\Bas}{{\it Basic}}
\newcommand{\Popper}{\mbox{\em Pop\/}}
\renewcommand{\T}{T}
\newcommand{\lab}{\mbox{\em label\/}}
\renewcommand{\aeq}{\approx}
\renewcommand{\naeq}{\!\!\approx}
\newcommand{\nsim}{\!\!\sim}
\newcommand{\stand}[1]{\mbox{\em st}\left (#1 \right )}
\renewcommand{\mid}{\, | \,}

\begin{document}

\title{Lexicographic probability, conditional probability, and
nonstandard probability%
\thanks{The work was supported in part by NSF under
grants IRI-96-25901, IIS-0090145, and CTC-0208535, by 
ONR under grant
N00014-02-1-0455, and by the DoD Multidisciplinary University Research
Initiative (MURI) program administered by the ONR under
grant N00014-01-1-0795. 
A preliminary version appeared in the Proceedings of the Eighth
Conference on Theoretical Aspects of Rationality and Knowledge, 2001
[Halpern 2001].
This version includes detailed proofs and more discussion
and more examples; in addition, the material in Section~\ref{sec:indep}
(on independence) is new. 
}}
\author{
Joseph Y.\ Halpern\\
Dept. Computer Science\\
Cornell University\\
Ithaca, NY 14853\\
halpern@cs.cornell.edu\\
http:/$\!$/www.cs.cornell.edu/home/halpern
}
\date{\today}
\maketitle
\begin{abstract}
The relationship between {\em Popper spaces\/} (conditional probability
spaces that satisfy some regularity conditions), 
lexicographic probability systems (LPS's) \cite{BBD1,BBD2}, and
nonstandard probability spaces (NPS's) is considered.  If countable
additivity is assumed, Popper spaces and a subclass of
LPS's are equivalent; without the assumption of countable additivity,
the equivalence no longer holds.  If the state space is finite, LPS's are
equivalent to NPS's.  However, if the state space is infinite, NPS's are
shown to be more general than LPS's.
\end{abstract}


\nocite{Hal26}
\section{Introduction}
Probability  is certainly the most commonly-used approach for
representing uncertainty and conditioning the standard way of updating
probabilities in the light of new information.  Unfortunately, there is a
well-known problem with conditioning:  Conditioning on events of measure
0 is not defined.  That makes it unclear how to proceed if an agent
learns something to which she initially assigned probability 0.
Although consideration of events of measure 0 may seem to be of little
practical interest, it turns out to play a critical role in game theory,
particularly in the analysis of strategic reasoning in extensive-form
games and in the analysis of weak dominance in normal-form games
(see, for example,
\cite{Bat96,BS02,BBD1,BBD2,BFK04,FT91a,Hammond94,Hammond99,KR97,KW82,Myerson86,Selten65,Selten75}). 
It also arises in the analysis of conditional 
statements by philosophers (see \cite{Adams66,McGee94}), and in dealing
with nonmonotonicity in Artificial Intelligence (see, for example,
\cite{LehmannMagidor}). 

There have been various attempts to deal with the problem of
conditioning on events of measure 0.  Perhaps the
best known 
involves \emph{conditional probability spaces} (CPS's).  The idea,
which goes back to Popper \citeyear{Popper34,Popper68} and
de Finetti 
\citeyear{Finetti36}, is to take as 
primitive not probability, but conditional probability.  If $\mu$ is a
conditional probability measure on a space $W$, then $\mu(V \mid U)$ may
still be undefined 
for some pairs $V$ and $U$, but it is also possible that $\mu(V \mid U)$ is
defined even if $\mu(U \mid W) = 0$.  A second approach, which goes back to at
least Robinson~\citeyear{Robinson73} and has been explored in
the economics literature \cite{Hammond94,Hammond99}, the AI literature
\cite{LehmannMagidor,Wilson95}, and the philosophy literature (see
\cite{McGee94} and the references therein) is to consider {\em
nonstandard probability spaces\/} (NPS's), where there are 
infinitesimals that can be used to model  
events that, intuitively, have infinitesimally small probability  yet
may still be learned or observed.  

There is a third approach to this problem, which  uses
sequences of probability measures to represent uncertainty.  The most
recent exemplar of this approach, which I focus on here,
are the {\em lexicographic probability systems\/} 
(LPS's)
of Blume, Brandenburger, and
Dekel \citeyear{BBD1,BBD2} (BBD from now on). However, the idea of using
a system of measures to represent uncertainty actually 
was explored as far back as the 1950s by R\'{e}nyi
\citeyear{Renyi56}
(see Section~\ref{related}).  
A {\em lexicographic
probability system\/} is a sequence 
$\<\mu_0,\mu_1, \ldots\>$ of probability measures.  Intuitively, the
first measure in the sequence, $\mu_0$, is the most important one,
followed by $\mu_1$, $\mu_2$, and so on.  
One way to understand LPS's is in terms of NPS's.
Roughly speaking, the
probability assigned to an event $U$ by a sequence such as
$\<\mu_0,\mu_1\>$ can be taken to be $\mu_0(U) + \epsilon\mu_1(U)$,
where $\epsilon$ is an infinitesimal.
Thus, even if the probability of $U$ according to $\mu_0$ is 0, $U$ 
still has a positive (although infinitesimal) probability if $\mu_1(U) > 0$.  

What is the precise relationship between these approaches? 
The relationship between LPS's and CPS's has been considered before.
For example, Hammond \citeyear{Hammond94} shows that
conditional probability spaces are equivalent to a subclass of LPS's
called \emph{lexicographic conditional probability spaces} (LCPS's) if
the state space 
is finite and it is possible 
to condition on any nonempty set.%
\footnote{Despite this isomorphism; it is not clear that conditional
probability spaces are \emph{equivalent} to LPS's.   It depends on
exactly what we mean by equivalence.  The same comment applies below
where the word ``equivalent'' is used.  See Section~\ref{discussion}
for further discussion.  I thank Geir Asheim for bringing this point to
my attention.} 
As shown by Spohn \citeyear{Spohn86},
Hammond's result can be extended to arbitrary countably additive {\em
Popper spaces}, where a Popper space is a conditional probability space
where the events on which conditioning is allowed satisfy certain
regularity conditions.
As I show, this result is depends critically on a number of assumptions.
In particular, it does not work without the assumption of countable
additivity, it requires that we extend LCPS's appropriately to the
infinite case, and it is sensitive to the choice of conditioning events.
For example, if we consider CPS's where
the conditioning events can be viewed as information sets, and so are 
are not closed under supsersets (this is
essentially the case considered by Battigalli and Sinischalchi
\citeyear{BS02}), then the result no longer holds.

Turning to the relationship between LPS's and NPS's,
I show that if the state space is finite, then LPS's are in a sense
equivalent to NPS's.  More precisely, say that two measures of
uncertainty $\nu_1$ and $\nu_2$ (each of which can be either an LPS or
an NPS) are equivalent, denoted $\nu_1 \aeq \nu_2$, if they cannot be
distinguished by (real-valued) 
random variables; that is, for all random variables $X$ and $Y$, 
$E_{\nu_1}(X) \le E_{\nu_1}(Y)$ iff $E_{\nu_2}(X) \le E_{\nu_2}(Y)$
(where $E_\nu(X)$ denotes the expected value of $X$ with respect to
$\nu$).
To the extent that we are interested in these representations
of uncertainty for decision making, then we should not try to
distinguish two representations that are equivalent.
I show that, in finite spaces, there is a straightforward bijection
between $\aeq$-equivalence classes of LPS's and NPS's.    This 
equivalence breaks down if the state space is infinite; in this case,
NPS's are strictly more general than LPS's
(whether or not countable additivity is assumed).  

Finally, I consider the relationship between Popper spaces and NPS's,
and show that NPS's are more general.
(The theorem I prove is a generalization of one proved 
by McGee \citeyear{McGee94}, but my interpretation of it is quite
different; see Section~\ref{PopperNPS}.)

These results give some useful insight into independence of random
variables.  There have been a number of alternative notions of
independence considered in the literature of extended probability spaces
(i.e., approaches that deal with the problem of conditioning on sets of
measure 0):  BBD considered three; Kohlberg and Reny \citeyear{KR97}
considered two others.   It turns out that these notions are perhaps
best understood in the context of NPS's; I describe and compare them
here.  

Many of the new results in this paper involve infinite spaces.
Given that most games studied by game theorists are finite, it is fair
to ask whether these results have any significance for game theory.
I believe they do.  Even if the underlying game is finite, the set of
types is infinite.  Epistemic characterizations of solution concepts
often make use of \emph{complete} type spaces,
which include every possible type of every
player, where a type determines an (extended) probability over the strategies
and types of the other players; this must be an infinite space.
For example, Battigalli and Siniscalchi \citeyear{BS02} use a
complete type space where the uncertainty is represented by cps's to
give an epistemic characterization of extensive-form rationalizability and
backward induction, while Brandenburger,  Friedenberg, and Keisler
\citeyear{BFK04} use a complete type space where the uncertainty is
represented by LPS's to get a characterization of weak dominance in
normal-form games. 
As the results of this paper show, the set of types depends 
to some extent 
on the notion of extended probability used.
Similarly, a number of characterizations of solution concepts depend on
independence (see, for example, \cite{Bat96,KR97,BS99a}).  Again, the results
of this paper show that these notions can be somewhat sensitive to
exactly how uncertainty is represented,
even with a finite state space.
While I do not present any new
game-theoretic results here, I believe that the characterizations I have
provided may be useful both in terms of defending particular choices of
representation used and suggesting new solution concepts.

The remainder of the paper is organized as follows.  In the next
section, I review all the relevant definitions for the three
representations of uncertainty considered here.  Section~\ref{FCP}
considers the relationship between Popper spaces and
LPS's.  Section~\ref{LPSNPS} considers the relationship between 
LPS's and NPS's.  Finally,
Section~\ref{PopperNPS} considers the relationship between Popper spaces
and NPS's.  In Section~\ref{sec:indep} I consider what these results
have to say about independence.  I conclude with
some discussion in Section~\ref{discussion}.

\section{Conditional, lexicographic, and nonstandard probability spaces}
In this section I briefly review the three approaches to representing
likelihood discussed in the introduction.

\subsection{Popper spaces}\label{cpsdef}

A {\em conditional probability measure\/} takes {\em pairs\/} $U, V$ of subsets
as arguments; $\mu(V,U)$ is generally written $\mu(V \mid U)$ to stress the
conditioning aspects. The first argument comes from some
algebra $\F$ of 
subsets of a space $W$; if $W$ is infinite, $\F$ is often taken to be a
$\sigma$-algebra.  (Recall that an algebra of subsets of $W$ is a set of
subsets containing $W$ and closed under union and complementation.  A
$\sigma$-algebra is an algebra that is closed under union countable.)
The second argument comes from a set $\F'$ of conditioning
events, that is, that is, events on which conditioning is allowed.
One natural choice is to take $\F'$ to be $\F - \emptyset$.  But it may
be reasonable to consider other restrictions on $\F'$.  For example,
Battigalli and Sinischalchi \citeyear{BS02} take $\F'$ to consist of the
information sets in a game, since they are interested only in agents who
update their beliefs conditional on getting some information.
The question is what
constraints, if any, should be placed on $\F'$.   
For most of this paper, I focus on \emph{Popper spaces} (named after
Karl Popper), defined next, where the set $\F'$ satisfies four arguably
reasonable requirements, but I occasionally consider other requirements
(see Section~\ref{sec:BS}).

\commentout{
\dfn A {\em Popper
algebra\/} over $W$ is a set $\F \times \F'$ of subsets of $W \times W$
such that (a) $\F$ is an algebra over $W$, (b) $\F'$ is a nonempty
subset of $\F$ (not necessarily an algebra over $W$) that does not
contain $\emptyset$, and
(c) $\F'$ is closed under supersets in $\F$, in that if
$V \in \F'$, $V \subseteq V'$, and $V' \in \F$, then $V' \in \F'$.
(Popper algebras are named after Karl Popper.) 
\edfn

While I have called these requirements ``minimal'', note that if $\F'$
is taken to consist of information sets, then it is not closed under
supersets in $\F$.  I return to this issue in Section~\ref{sec:BS}.
}

\dfn\label{dfn.condprob} A {\em conditional probability space (cps) over
$(W,\F)$\/} is a tuple
$(W,\F,\F',\mu)$ such that 
$\F$ is an algebra over $W$, $\F'$ is a set of subsets of $W$ 
(not necessarily an algebra over $W$) that does not
contain $\emptyset$, and
$\mu: \F \times \F' \rightarrow [0,1]$ satisfies the
following conditions:
\begin{itemize}
\item[CP1.] $\mu(U \mid U) = 1$ if $U \in\F'$.
\item[CP2.] $\mu(V_1 \union V_2 \mid U) = \mu(V_1 \mid U) + \mu(V_2 \mid U)$ if $V_1
\inter V_2 = \emptyset$, $U \in \F'$, and $V_1, V_2 \in \F$.
\item[CP3.] $\mu(V \mid U) = \mu(V \mid X) \times \mu(X \mid U)$ if $V \subseteq X
\subseteq U$, $U, X \in \F'$, $V \in \F$.
\end{itemize}
Note that it follows from CP1 and CP2 that $\mu(\cdot \mid U)$ is a
probability measure on $(W,\F)$ (and, in particular, that $\mu(\emptyset
\mid U) = 0$) for each $U \in \F'$.
A {\em Popper space over $(W,\F)$\/} is a conditional probability space
$(W,\F,\F',\mu)$ 
that satisfies 
three additional conditions: (a) $\F' \subseteq \F$, (b) 
$\F'$ is closed under supersets in $\F$, in that if
$V \in \F'$, $V \subseteq V'$, and $V' \in \F$, then $V' \in \F'$, and 
(c) if $U \in \F'$ and $\mu(V \mid U) \ne 0$ then $V \inter U \in \F'$.  
If $\F$ is a $\sigma$-algebra and $\mu$ is countably additive
(that is, if $\mu(\union V_i \mid U) = \sum_{i = 1}^\infty \mu(V_i \mid U)$ if the
$V_i$'s are pairwise disjoint elements of $\F$ and $U \in \F'$), then the
Popper space is said to be {\em countably additive}.
Let $\Popper(W,\F)$ denote the set of Popper spaces over $(W,\F)$. 
If $\F$ is a $\sigma$-algebra, I use a superscript $c$ to
denote the restriction to countably additive Popper spaces, so
$\Popper^c(W,\F)$ denotes the set of
countably additive Popper spaces over $(W,\F)$.
The probability measure $\mu$ in a Popper space is
called a {\em Popper measure}.  
\edfn
The last regularity condition on $\F'$ required in a Popper space
corresponds to the observation that for an unconditional
probability measure $\mu$, if $\mu(V \mid U) \ne 0$ then $\mu(V \inter U)
\ne 0$, so conditioning on $V \inter U$ should be defined.
Note that, since this regularity condition depends on the Popper
measure, it may well be the case that $(W,\F,\F',\mu)$ and
$(W,\F,\F',\nu)$ are both cps's over $(W,\F)$, but only the former is a
Popper space over $(W,\F)$.

Popper \citeyear{Popper34,Popper68}\index{Popper, K.~R.} 
and de Finetti \citeyear{Finetti36} were the first to
formally consider conditional probability as the basic notion, although
as R\'{e}nyi \citeyear{Renyi64}\index{R\'{e}nyi, A.} points out, the
idea of taking conditional probability as primitive seems to go back as
far as Keynes \citeyear{Keynes}. 
CP1--3 are essentially due to R\'{e}nyi \citeyear{Renyi55}.
Van Fraassen \citeyear{vF76} defined what I have called Popper measures;
he called them Popper functions, reserving the name Popper measure for
what I am calling a countably additive Popper measure.
Starting from the work of de Finetti, there has been a general study of 
\emph{coherent conditional probabilities}.  A coherent conditional
probability is essentially a 
cps that is not necessarily a Popper space, since it is
defined 
on a set $\F \times \F'$ 
where $\F'$ does not have to be a subset of $\F$); see, for example,
\cite{CS02} and the references therein.
Hammond \citeyear{Hammond94} discusses the use of conditional
probability spaces in philosophy and game theory, and provides an
extensive list of references.


\subsection{Lexicographic probability spaces}\label{LPSdef}

\dfn A {\em lexicographic probability space (LPS) (of length
$\alpha$) over $(W,\F)$\/} is a tuple 
$(W,\F,\vecmu)$ where, as before, $W$ is a set of possible worlds and
$\F$ is an algebra over $W$, and $\vecmu$ is a sequence of finitely additive
probability measures on $(W,\F)$ indexed by ordinals $< \alpha$.
(Technically, $\vecmu$ is a function from the ordinals less
than $\alpha$ to probability measures on $(W,\F)$.)
I typically write $\vecmu$ as $(\mu_0, \mu_1, \ldots)$ or as
$(\mu_\beta: \beta < \alpha)$.
If $\F$ is a $\sigma$-algebra and each of the probability measures in
$\vecmu$ is countably additive, then $\vecmu$ is a {\em countably
additive LPS}.
Let $\LPS(W,\F)$ denote the set of LPS's over $(W,\F)$. 
Again, if $\F$ is a $\sigma$-algebra, a superscript $c$ is used to
denote countable additivity, so $\LPS^c(W,\F)$ denote the set of
countably additive LPS's over $(W,\F)$.
When $(W,\F)$ are understood, I often refer to $\vecmu$ as
the LPS.  
I write $\vecmu(U) > 0$ if $\mu_\beta(U) > 0$ for some $\beta$.
\edfn

There is a sense in which $\LPS(W,\F)$ can capture a richer set of
preferences than $\Popper(W,\F)$, 
even if we restrict to finite 
spaces $W$ (so that countable additivity is not an issue).  
For example, suppose that $W = \{w_1,w_2\}$, $\mu_0(w_1) = \mu_0(w_2) =
1/2$, and $\mu_1(w_1) = 1$.  The LPS $\vecmu = (\mu_0,\mu_1)$  can be
thought of describing the situation where $w_1$ is very slightly more
likely than $w_2$.  Thus, for example, if $X_i$ is a bet that pays off 1
in state $w_i$ and 0 in state $w_{3-i}$, then according to $\vecmu$,
$X_1$ should be (slightly) prefereed to $X_2$, but for all $r > 1$,
$rX_2$ is preferred to $X_1$.  There is no CPS on $\{w_1,w_2\}$ that
leads to these preferences 

Note that, in this example, the support of $\mu_2$ is a subset of that
of $\mu_1$.  To obtain a bijection between LPS's and CPS's, we cannot
allow much overlap between the supports of the measures that make an
LPS.  What counts as ``much overlap'' turns out to be a somewhat subtle.
One way to formalize it was proposed by BBD.  They defined a {\em  
lexicographic conditional probability space (LCPS)\/} to be an LPS such
that,
roughly speaking,
the probability measures in the sequence have disjoint supports;
more precisely, there exist sets $U_\beta \in \F$ such that $\mu_\beta(U_\beta)
= 1$ and the sets $U_\beta$ are pairwise disjoint for $\beta < \alpha$.
One motivation for considering disjoint sets is to consider an agent who
has a sequence of hypotheses $(h_0, h_1, \ldots)$ regarding how the
world works.  If the primary hyothesis $h_0$ is discarded, then the
agent judges events according to $h_1$; if $h_1$ is discarded, then the
agent uses $h_2$, and so on.  Associated with hypothesis $h_\beta$ is
the probability measure $\mu_\beta$.  What would cause $h_\beta$ to be
discarded is observing an event $U$ such that $\mu_\beta(U) = 0$.
The set $U_\beta$ is the support of the hypothesis $h_\beta$.  In some
cases, it seems reasonable to think of the supports of these hypotheses
as disjoint.  This leads to LCPS's.  

BBD considered only finite spaces.  When we move to infinite spaces,
requiring disjointness of the supports of hypotheses may be too strong.  
Brandenburger, Friedenberg, and Keisler \citeyear{BFK04} consider
finite-length LPS's $\vecmu$ that satisfy the property that
there exist sets $U_\beta$ (not necessarily disjoint) such that
$\mu_\beta(U_\beta) = 1$ and $\mu_\beta(U_\gamma) = 0$ for $\gamma \ne
\beta$.  Call such an LPS an MSLPS (for \emph{mutually singular LPS}).
Let a {\em structured LPS (SLPS)\/} be an LPS $\vecmu$ such that there
exist sets $U_\beta \in \F$ such that $\mu_\beta(U_\beta) = 1$ and
$\mu_\beta(U_\gamma) = 0$ for $\gamma > \beta$. 
Thus, in an SLPS, later hypotheses are given probability 0 according to
the probability measure induced by earlier hypotheses, but earlier 
hypotheses do not necessarily get probability 0 according the later
hypotheses.  (Spohn~\citeyear{Spohn86} also considered SLPS's; he called
them {\em dimensionally well-ordered families of probability measures}.)  
Clearly every LCPS is an MSLPS, and every MSLPS is an SLPS.  If $\alpha$
is countable and we require countable additivity (or if $\alpha$ is
finite) then the notions are easily seen to coincide.  Given an SLPS
$\vecmu$ with associated sets $U_\beta, \beta < 
\alpha$, define $U_\beta' = U_\beta - (\union_{\gamma > \beta} U_\gamma)$.  
The sets $U_\beta'$ are
clearly pairwise disjoint elements of $\F$, and 
$\mu_\beta(U_\beta') = 1$.
However, in general, LCPS's are a strict subset of MSLPS's, and MSLPS's 
are a strict subsets of SLPS's, as the following two examples show.

\commentout{
To understand the motivation for SLPS's, consider an agent with a
sequence of hypotheses (modeled as probability distributions).  The first
hypothesis, modeled by $\mu_0$, is used as long as it is not
controverted by  evidence.  If an event $E$ is discovered that shows
that the first hypothesis must be wrong (i.e., $\mu_0(E) = 0)$, then
next hypothesis that gives $E$ positive measure is used.  With this
intuition, it seems reasonable that if $i > j$, the set of states where
$\mu_j$ is used, namely, $U_j$, should be a set that is given measure 0
by all $\mu_i$ with $i < j$; hypothesis $j$ should not be used unless
all higher-ranking hypotheses have been discarded.
}


\xam\label{SLPSxam}  Consider a well-ordering of the interval $[0,1]$,
that is, 
a bijection
from $[0,1]$ to an initial segment of the ordinals. 
Suppose that this initial segment of the ordinals has length $\alpha$.
Let $([0,1],\F,\vecmu)$ be an LPS of length $\alpha$ where $\F$ consists
of the Borel subsets of $[0,1]$.
Let $\mu_0$ be the standard Borel measure on $[0,1]$,
and let $\mu_\beta$ be the measure that gives probability 1 to
$r_\beta$, the $\beta$th real in the well-ordering.  This clearly gives
an SLPS, since 
we can take $U_0 = [0,1]$ and $U_\beta = \{r_\beta\}$ for $0
< \beta < \alpha$; note that $\mu_\alpha(U_\beta) = 0$ for $\beta > \alpha$.
However, this SLPS is not equivalent to any MSLPS (and hence not to any
LCPS); there 
is no
set $U_0'$ such that $\mu_0(U_0') = 1$ and $U_0'$ is disjoint from
$r_\beta$ for all $\beta$ with $0 < \beta < \alpha$.
\exam

\xam\label{MSLPSxam}  Suppose that $W = [0,1] \times [0,1]$.  Again,
consider a well-ordering on $[0,1]$.  Using the notation of
Example~\ref{SLPSxam}, define $U_{0,\beta} = r_{\beta} \times [0,1]$ and
$U_{1,\beta} = [0,1] \times \{r_\beta\}$.  Define $\mu_{i,\beta}$ to be
the Borel measure on $U_{i,\beta}$.  Consider the LPS $(\mu_{0,0},
\mu_{0,1}, \ldots, \mu_{1,0}, \mu_{1,1}, \ldots)$.  Clearly this is an
MSLPS, but not an LCPS.  \exam

The difference between LCPS's, MSLPS's, and SLPS's does not arise in the work
of BBD, since they consider only finite
sequences of measures.  The restriction to finite sequences, in turn, is
due to their restriction to finite sets $W$ of possible worlds.  
Clearly, if $W$ is finite, then all LCPS's over $W$ must have length $\le
|W|$, since the measures in an LCPS have disjoint supports.  Here it
will play a more significant role.

We can put an obvious lexicographic order $<_L$ on sequences $(x_0, x_1,
\ldots)$ of numbers in $[0,1]$ of length $\alpha$: $(x_0, x_1, \ldots)
<_L (y_0, y_1, \ldots)$ if there exists $\beta < \alpha$ such that
$x_\beta < y_\beta$ and $x_\gamma
= y_\gamma$ for all $\gamma < \beta$.  That is, we
compare two sequences by comparing their components at the first place
they differ.  (Even if $\alpha$ is infinite, because we are dealing with
ordinals, there will be a least ordinal at which the sequences differ if
they differ at all.)  This lexicographic order will be used 
to define decision rules.  

BBD define conditioning in LPS's as follows.  Given $\vecmu$ and $U \in
\F$ such that $\vecmu(U) > 0$, let $\vecmu|U =
(\mu_{k_0}(\cdot \mid U), \mu_{k_1}(\cdot \mid U), \ldots )$, where $(k_0,
k_1, \ldots)$ is the subsequence of all indices for which the
probability of $U$ is positive.  
Formally, $k_0 = \min\{k: \mu_k(U) > 0\}$
and for an arbitrary ordinal $\beta > 0$, if $\mu_{k_\gamma}$ has been
defined for all $\gamma < \beta$ and there exists a measure $\mu_{\delta}$ in
$\vecmu$ such that $\mu_{\delta}(U) > 0$ and $\delta > k_\gamma$ for all
$\gamma < \beta$, then $k_\beta = \min\{\delta: \mu_{\delta}(U) > 0, \,
\delta > k_\gamma \mbox{ for all } \gamma < \beta\}$.  
Note that
$\vecmu|U$ is undefined if $\vecmu(U) = 0$.

\subsection{Nonstandard probability spaces}\label{NPSdef}

It is well known that there exist {\em non-Archimedean fields}---fields
that include the real  numbers as a subfield but also have 
{\em infinitesimals\/}, numbers that are positive but still less than
any positive real number.   The smallest such non-Archimedean field,
commonly denoted $\IR(\epsilon)$, is the smallest field generated by
adding to the reals a single infinitesimal $\epsilon$.%
\footnote{The construction of $\IR(\epsilon)$ apparently goes back to
Robinson \citeyear{Robinson73}.  It is reviewed by 
Hammond \citeyear{Hammond94,Hammond99} and Wilson
\citeyear{Wilson95} (who calls $\IR(\epsilon)$ the {\em extended
reals\/}).}
We can further restrict to non-Archimedean fields that are
\emph{elementary extensions} of the standard reals: they 
agree with the
standard reals on all properties that can be expressed in a first-order
language with a predicate $N$ representing the natural numbers.
For most of this paper, I use only the following properties 
of non-Archimedean fields:
\begin{enumerate}
\item If $\IR^*$ is a non-Archimedean field, then for all $b \in \IR^*$
such that $-r < b < r$ for some standard real $r > 0$, 
there is a unique closest real number $a$ such that $|a - b|$ is an
infinitesimal.  (Formally, $a$ is the inf of the set of real numbers
that are at least as large as $b$.)  Let $\stand{b}$ denote the closest
standard real to $b$; $\stand{b}$ is sometimes read ``the standard
part of $b$''.
\item If $\stand{\epsilon/\epsilon'} =0$, then $a \epsilon < \epsilon'$
for all positive standard real numbers $a$.  (If $a \epsilon$ were
greater than $\epsilon'$, then $\epsilon/\epsilon'$ would be greater
than $1/a$, 
contradicting the assumption that $\stand{\epsilon/\epsilon'} = 0$.)
\end{enumerate}

Given a non-Archimedean field $\IR^*$, a  {\em
nonstandard probability space (NPS) over $(W,\F)$ (with range $\IR^*$)\/} is
a tuple $(W,\F,\mu)$, where $W$ is a set of possible worlds,  $\F$ is an
algebra 
of subsets of $W$, and $\mu$ assigns to sets in $\F$ 
a nonnegative element of
$\IR^*$ such that $\mu(W) = 1$ and $\mu(U \union V) = \mu(U) + \mu(V)$ if
$U$ and $V$ are disjoint.%
\footnote{Note that, unlike Hammond \citeyear{Hammond94,Hammond99},
I do not restrict the range of probability measures to consist of 
ratios of polynomials in $\epsilon$ with nonnegative coefficients.}

If $W$ is infinite, we may also require that
$\F$ be a $\sigma$-algebra and that $\mu$ be countably additive.  
(There are some subtleties involved with countable additivity in
nonstandard probability spaces; see Section~\ref{countableadditivity}.)

\section{Relating Popper Spaces to (S)LPS's}\label{FCP}

In this section, I consider a mapping $\FCP$ from SLPS's over
$(W,\F)$ to Popper spaces over $(W,\F)$, for each fixed $W$ and $\F$,
and show that, in many cases of interest, $\FCP$ is a bijection.
Given an SLPS $(W,\F,\vecmu)$ of length $\alpha$,
consider the cps $(W,\F,\F',\mu)$ such that $\F' = \union_{\beta <
\alpha} \{V \in \F: \mu_\beta(V)
> 0 \}$.  For $V \in \F'$, let $\beta_V$
be the smallest index such $\mu_{\beta_V}(V) > 0$.  Define $\mu(U \mid V) =
\mu_{\beta_V}(U \mid V)$.  I leave it to the reader to check that
$(W,\F,\F',\mu)$ is a Popper space.

There are many bijections between two spaces.
Why is $\FCP$ of interest?
Suppose that $\FCP(W,\F,\vecmu) = (W,\F,\F',\mu)$.  It is easy
to check that the following two important properties hold:
\begin{enumerate}
\item $\F'$ consists precisely of those events for which conditioning in
the LPS is defined; that is, $\F' = \{U: 
\vecmu(U) > 0\}$.
\item For $U \in \F'$, $\mu(\cdot \mid U) = \mu'(\cdot \mid U)$, where
$\mu'$ is the first probability measure in the sequence $\vecmu|U$.
That is, the 
Popper measure agrees with the most significant probability measure
in the conditional LPS given $U$.  Given that an LPS assigns to an event
$U$ a sequence of numbers and a Popper measure assigns to $U$ just a
single number, this is clearly the best single number to take.
\end{enumerate}
It is clear that these two properties in fact characterize $\FCP$.
Thus, $\FCP$ preserves the events on which conditioning is possible and
the most significant term in the lexicographic probability.

\subsection{The finite case}
It is useful to separate the analysis of $\FCP$ into two cases, depending
on whether or not the state space is finite.  I consider the finite case
first.  

BBD claim without proof that $\FCP$ is a bijection
from LCPS's to  
conditional probability spaces.  They work in finite spaces $W$ (so that
LCPS's are equivalent to SLPS's) and restrict
attention to LPS's where $\F
= 2^W$ and $\F' = 2^W - \{\emptyset\}$ (so that conditioning is defined for
all nonempty sets).  Since $\F' = 2^W - \{\emptyset\}$, the cps's they
consider are all Popper spaces. 
Hammond \citeyear{Hammond94} provides a careful proof of this result,
under the restrictions considered by BBD.  
I generalize Hammond's result by considering 
finite Popper spaces
with arbitrary conditioning events.
No new conceptual issues arise in doing this extension; I
include it here only to be able to contrast it with the other
results. 

Let $\SLPS(W,\F)$ denote the set of LPS's over $(W,\F)$; let
$\SLPS(W,\F,\F')$ denote the set of LPS's $(W,\F,\vecmu)$ such that
$\vecmu(U) > 0$ for all $U \in \F'$ (i.e., $\mu_\beta(U) > 0$ for some
$\beta$); as usual, I use a superscript $c$ to denote countable
additivity, so, for example, $\SLPS^c(W,\F)$ denotes the set of
countably additive SLPS's over $(W,\F)$.
Let $\Popper(W,\F,\F')$ denote the set of Popper spaces of the form
$(W,\F,\F')$ and let  $\Popper^c(W,\F,\F')$
denote the set of Popper spaces of the form
$(W,\F,\F',\mu)$ where $\mu$ is countably additive.

\thm\label{FCPfin} 
If $W$ is finite, then
$\FCP$ is a bijection from $\SLPS(W,\F,\F')$ to $\Popper(W,\F,\F')$.  \ethm

\prf It is immediate from the definition that if $(W,\F,\vecmu) \in
\SLPS(W,\F,\F')$, then $\FCP(W,\F,\vecmu) \in \Popper(W,\F,\F')$.  It is
also straightforward to show that $\FCP$ is an injection (see the
appendix for details).  The work comes in showing that $\FCP$ is a
surjection (or, equivalently, in constructing an inverse to $\FCP$).
I sketch the main ideas of the argument here, leaving details to the
appendix.  

Given $\mu \in \Popper(W,\F,\F')$, the idea is to choose $k \le |W|$ and $k$ 
disjoint sets $U_0, \ldots, U_k \in \F'$ appropriately such that $\mu_j
= \mu \mid U_j$ for $j= 0, \ldots, k$ (i.e., $\mu_j(V) = \mu(V \mid
U_j)$) amd $\FCP(W,\F,\vecmu) = \mu$.  Since the sets $U_0, \ldots, U_k$
are disjoint, $\vecmu$ must be an SLPS.  The difficulty lies in choosing
$U_0, \ldots, U_k$ so that $\vecmu(U) > 0$ iff $U \in \F'$.  This is
done as follows.  Let $U_0$ be the smallest set $U \in \F$ such that
$\mu(U) = 1$.   
Since $W$ is finite, there is such a smallest set; it is simply the
intersection of all sets $U$ such that $\mu(U \mid W) = 1$.  Since $\mu(U_0
\mid W) > 0$, it follows that $U_0 \in \F'$.  If $\overline{U}_0 \notin
\F'$. then (because $\F'$ is closed under supersets in $\F$), no
subset of $\overline{U}_0$ is in $\F'$.  If $\overline{U}_0 \in
\F'$, let $U_1$ be the smallest set in 
$\F$ such that $\mu(U_1 \mid \overline{U}_0) = 1$.  Note that $U_1
\subseteq \overline{U}_0$ and that $U_1 \in \F'$.
Continuing in this way, it is
clear that there exists a $k \ge 0$ and a sequence of pairwise disjoint
sets $U_0, U_1, \ldots, U_k$ such that (1) $U_i \in \F'$ for $i = 0,
\ldots, k$,  (2) for $i < k$, $\overline{U_0 \union \ldots \union U_i}
\in \F'$ and $U_{i+1}$ is the smallest subset of $\F$ such that 
$\mu(U_{i+1} \mid \overline{U_0 \union \ldots \union U_i}) = 1$, and (3)
$\overline{U_0 \union \ldots \union U_k} \notin \F'$. 
Condition (2) guarantees that $U_{i+1}$ is a subset of 
$\overline{U_0 \union \ldots \union U_i}$, so the $U_i$'s are pairwise
disjoint.  Define the LPS $\vecmu = (\mu_1, \ldots, \mu_k)$ by taking
$\mu_i(V) = \mu(V \mid U_i)$.  Clearly the support of $\mu_i$ is $U_i$, so
this is an LCPS (and hence an SLPS).  \eprf


\cor\label{FCPfin1} 
If $W$ is finite, then $\FCP$ is a bijection
from $\SLPS(W,\F)$ to $\Popper(W,\F)$.  \ecor



\subsection{The infinite case}

The case where the state space $W$ is infinite is not considered
by either BBD or Hammond.  It presents some interesting subtleties.

It is easy to see that $\FCP$ is an injection from 
SLPS's to Popper spaces.  However, 
as the following two examples show, if we do not require countable
additivity,  then it is not a bijection.

\xam\label{counter1}  (This example is essentially due to Robert
Stalnaker [private communication, 2000].)  Let $W = \IN$, the natural
numbers, let $\F$ consist of the finite and cofinite subsets of $\IN$
(recall that a cofinite set is the complement of a finite set),
and let $\F' = \F - \{\emptyset\}$.  If $U$ is cofinite,
take $\mu^1(V \mid U)$  to be 1 if $V$ is cofinite and 0 if $V$ is finite.
If $U$ is finite, define $\mu^1(V \mid U) = |V \inter
U|/|U|$.  I leave it to the reader to check that $(\IN,\F,\F',\mu^1)$ is a
Popper space.  Note that $\mu^1$ is not countably additive (since
$\mu^1(\{i\} \mid \IN) = 0$ for all $i$, although $\mu^1(\IN \mid \IN) =
1$). 
Suppose that there were some LPS $(\IN,\F,\vecmu)$ which was
mapped by $\FCP$ to this Popper space.  Then it is easy to check that
if $\mu_i$ is the first measure in $\vecmu$ such that $\mu_i(U) > 0$ for
some finite set $U$, then $\mu_i(U') > 0$ for all nonempty finite sets $U'$.
To see this, note that for any nonempty finite set $U'$, since
$\mu_i(U) > 0$, it follows that $\mu_i(U \union U') > 0$.  Since $U
\union U'$ is finite, it must be the case that $\mu_i$ is the first
measure in $\vecmu$ such that $\mu_i(U \union U') > 0$.  Thus, by
definition, $\mu^1(U' \mid U \union U') = \mu_i(U' \mid U \union U')$.  Since
$\mu^1(U' \mid U \union U') > 0$, it follows that $\mu_i(U') > 0$.  
Thus, $\mu_i(U') > 0$ for all nonempty finite sets $U'$.

It is also easy to see that $\mu_i(U)$ must be proportional
to $|U|$ for all finite sets $U$.  To show this, it clearly suffices to show
that $\mu_i(n) = \mu_i(0)$ for all $n \in \IN$.  But this is immediate
from the observation that 
$$\mu_i(\{0\} \mid \{0, n \}) = \mu^1(\{0\} \mid \{0, n \}) =
|\{0\}|/|\{0,n\}| = \frac{1}{2}.$$
But there is no probability measure $\mu_i$ on the natural
numbers such that $\mu_i(n) = \mu_i(0) > 0$ for all $n \ge 0$.  
For if $\mu_i(0) > 1/N$, then $\mu_i(\{0, \ldots, N-1\}) > 1$, a
contradiction.  
(See Example~\ref{counter3} for further discussion of this setup.)
\exam

\xam\label{counter2}  Again, let $W = \IN$, 
let $\F$ consist of the finite and cofinite subsets of $\IN$,
and let $\F' = \F - \{\emptyset\}$.  As with $\mu^1$, 
if $U$ is cofinite,
take $\mu^2(V \mid U)$  to be 1 if $V$ is cofinite and 0 if $V$ is finite.
However, now, if $U$ is finite, define $\mu^2(V \mid U) =
1$ if $\max(V \inter U) = \max U$, and $\mu^2(V \mid U) = 0$ otherwise.
Intuitively, if $n > n'$, then $n$ is infinitely more probable than $n'$
according to $\mu^2$.
Again, I leave it to the reader to check that $(\IN,\F,\F',\mu^2)$ is a
Popper space.  Suppose there were some LPS $(\IN,\F,\vecmu)$ which was
mapped by $\FCP$ to this Popper space.  Then it is easy to check that
if $\mu_n$ is the first measure in $\vecmu$ such that $\mu_n(\{n\}) > 0$, then
$\mu_n$ comes before $\mu_{n'}$ in $\vecmu$ if $n > n'$.  However, since
$\vecmu$ is well-founded, this is impossible.
\exam

As the following theorem, 
originally
proved by Spohn \citeyear{Spohn86}, shows,
there are no such counterexamples if we 
restrict to countably additive SLPS's and countably additive Popper spaces.

\thm\label{infiso} {\rm \cite{Spohn86}}
For all $W$, the map $\FCP$ is a bijection from 
$\SLPS^c(W,\F,\F')$
to $\Popper^c(W,\F,\F')$.  \ethm

\prf Again, the difficulty comes in showing that $\FCP$ is onto.  Given 
a Popper space $(W,\F,\F',\mu)$, I again construct sets $U_0, U_1,
\ldots$ and an LPS $\vecmu$ such that $\mu_\beta(V)=\mu(V \mid
U_\beta)$, and show that $\FCP(W,\F,\vecmu) = (W,\F,\F',\mu)$.  
However, now a completely different construction is required; the
earlier inductive construction of the 
sequence $U_0, \ldots, U_k$ no longer works.  The problem already arises
in the construction of $U_0$.  There may no longer be a smallest set
$U_0$ such that $\mu(U_0) = 1$.  Consider, for example, the interval
$[0,1]$ with Borel measure.  There is clearly no smallest subset $U$ of
$[0,1]$ such that $\mu(U) = 1$.  The details can be found in the appendix.
\eprf

\cor\label{infiso1} For all $W$, the map $\FCP$ is a bijection from 
$\SLPS^c(W,\F)$
to $\Popper^c(W,\F)$.
\ecor

It is important in Corollary~\ref{infiso1} that we consider SLPS's and not
MSLPS's or LCPS's.  $\FCP$ is in fact not a bijection from MSLPS's or
LCPS's to Popper 
spaces.  

\xam\label{SLPSxam2} Consider the Popper space $([0,1],\F,\F',\mu)$ 
which is the image under $\FCP$ of the SLPS constructed in
Example~\ref{SLPSxam}.  It is easy to see that this Popper space cannot
be the image under $\FCP$ of some MSPLS (and hence not of some LCPS
either). \exam 

\subsection{Treelike CPS's}\label{sec:BS}

One of the requirements in a Popper space is that 
$\F'$ be closed under supersets in $\F$.  If we
think of $\F'$ as consisting of all sets on which conditioning is
possible, this makes sense; if we can condition on a set $U$, we should
be able to consider on a superset $V$ of $U$.  But if we think of $\F'$
as representing all the possible evidence that can be obtained (and
thus, the set of events on which an agent must be be able to
condition, so as to update her beliefs), there is no reason that $\F'$
should be closed under supersets; nor, for that matter, is it
necessarily the case that if $U \in \F'$ and $\mu(V \mid U) \ne 0$, then
$V \inter U \in \F'$.   
In general, a cps where $\F'$ does not have these properties
cannot be represented by an LPS, as the following
example shows.

\xam\label{xam:noPopper}  Let $W = \{w_1, w_2, w_3, w_4\}$, let $\F$ consist
of all subsets of $W$, and let $\F'$ consist of all the 2-element
subsets of $W$.  
Clearly 
$\F'$ is not closed under supersets.  Define $\mu$ on $\F \times \F'$
such that $\mu(w_1 \mid \{w_1,w_3\}) = \mu(w_4 \mid 
\{w_2,w_4\}) = 1/3$, and  $\mu(w_1 \mid \{w_1,w_2\}) = \mu(w_4 \mid
\{w_3,w_4\}) = 
1/2$, and CP1 and CP2 hold.  This is easily seen to determine $\mu$.
Moreover, $\mu$ vaciously satisfies CP3, since there do not exist
distinct sets $U$ and $X$ in $\F'$ such that $U \subseteq X$.  It is
easy to show that there is no unconditional probability $\mu^*$ on $W$
such that $\mu^*(U \mid V) = \mu(U \mid V)$ for all pairs $(U,V) \in \F
\times \F'$ such that $\mu^*(V) > 0$ (where, for $\mu^*$, the
conditional probability is defined in the standard way).%
\footnote{This example is closely related to examples of conditional
probabilities for which there is no common prior; see, for example,
\cite[Example 2.2]{Hal21}.}  It easily follows that there is no LPS
$\vecmu$ such that $\vecmu(U \mid V) = \mu(U \mid V)$ for all $(U,V) \in
\F \times \F'$ (since otherwise $\mu_0$ would agree with $\mu$ on all
pairs $(U,V) \in  \F \times \F'$ such that $\mu(V) > 0$). 
Had $\F'$ been closed under supersets, it would have included $W$.  It 
is easy to see that it is impossible to extend $\mu$ to $\F \times (\F'
\union \{W\})$ so that CP3 holds.
\exam

In the game-theory literature, Battigalli and Siniscalchi
\citeyear{BS02} use conditional probability measures to model players'
beliefs about other players' strategies in 
extensive-form games where agents have perfect recall.  The conditioning
events are essentially 
information sets;
which can be thought of as representing the possible evidence that an
agent can obtain in a agame.
Thus, the cps's they consider are not necessarily Popper spaces, for the
reasons described above.
Nevertheless, the conditioning events considered by Battigalli and
Sinischalchi  satisfy certain properties that
prevent an analogue of Example~\ref{xam:noPopper} from holding.
I now make this precise.

Formally, I assume that there is a one-to-one
correspondence between the sets in $\F'$ and the information sets of
some fixed player $i$.  For each set $U \in \F'$, there is a unique 
information set $I_U$ for player $i$ such that $U$ consists of all the
strategy profiles 
that reach $I_U$.  With this identification, it is immediate that we can
organize the sets in $\F'$ into a forest (i.e., a collection of trees),
with the same ``reachability'' structure as that of the information sets
in the game tree.  The topmost sets in the forest are the ones
corresponding to the topmost information
sets for player $i$ in the game tree.  There may be several such topmost
information sets if nature or some player $j$ other than $i$ makes the
first move in the game.  (That is why we have a forest, rather than a
tree.)  The immediate successors of a set $U$ are the sets of strategy
profiles corresponding to information sets for player $i$ reached
immediately after $I_U$. 
Because agents have perfect recall, the conditioning events $\F'$ have
the following properties:  
\begin{itemize}
\item[T1.] $\F'$ is countable.
\item[T2.] The elements of $\F'$ can be organized as a forest (i.e., a collection of
trees) where, for each $U \in \F'$, if there is an edge from $U$ to
some $U' \in \F'$, then $U' \subseteq U$, all the immediate successors
of  $U$ are 
disjoint, and $U$ is the union of its immediate successors.
\item[T3.] The topmost nodes in each tree of the forest form a
partition of $W$. 
\end{itemize}

Say that a set $\F'$ is \emph{treelike} if it satisfies T1--3.
It follows from T2 and T3 that, for any sets $U$ and $U'$ in a treelike
set $\F'$, either $U \subseteq U'$ (if $U$ is a descendant of $U'$
in some tree), $U' \subseteq  
U$ (if $U'$ is a descendant of $U$), or $U$ and $U'$ are disjoint (if
neither is a descendant of the other).  
If $\F'$ is treelike, let $\T^c(W,\F,\F')$ consist of all countably
additive cps's defined on 
$\F \times \F'$.  
I abuse notation
in the next result, viewing $\FCP$ as a mapping from 
$\SLPS^c(W,\F,\F')$ to $\T^c(W,\F,\F')$.

\pro\label{prop:BS} The map $\FCP$ is a surjection from 
$\SLPS^c(W,\F,\F')$ onto $\T^c(W,\F,\F')$.  \epro

Since $\F'$ is countable, every SLPS in $\SLPS^c(W,\F,\F')$ must have at
most countable length.  Thus, there is no distinction between SLPS's,
LCPS's, and MSPLS's in this case.  (Indeed, in the proof of
Proposition~\ref{prop:BS}, the LPS constructed to demonstrate the
surjection is an LCPS.)  Note that we cannot hope to get a bijection
here, even if $W$ is finite.  For example, suppose that $W = \{w_1,
w_2\}$, $\F = 2^W$, and  
$\F' = \{\{w_1\}, \{w_2\}\}$.  $\F'$ is clearly treelike, and there
is a unique cps $\mu$ on $(W,\F,\F')$.  $\FCP$ maps
every SLPS in $\SLPS(W,\F,F')$  to $\mu$, but is clearly not  a
bijection.  (This example also shows that we do not get a bijection by
considering MSLPS's or LCPS's either.)

\subsection{Related Work}\label{related}

It is interesting to contrast these results to those of R\'{e}nyi
\citeyear{Renyi56} and van Fraassen \citeyear{vF76}.  R\'{e}nyi considers
what he calls {\em dimensionally ordered\/} systems.  
A dimensionally ordered system over $(W,\F)$ has the form
$(W,\F,\F',\{\mu_i: i \in I\})$, where $\F$ is an algebra of
subsets of $W$, $\F'$ is a subset of $\F$ closed under finite unions
(but not necessarily closed under supersets in $\F$),
$I$ is a totally ordered set (but not necessarily well-founded, so it
may not, for example, have a first element) and $\mu_i$ is a measure on
$(W,\F)$ (not necessarily a probability measure) such that 
\begin{itemize}
\item for each $U \in \F'$, there is some $i \in I$ such that $0 <
\mu_i(U) < \infty$ (note that the measure of a set may, in general, be
$\infty$), 
\item if $\mu_i(U) < \infty$ and $j < i$, then $\mu_j(U) = 0$.
\end{itemize}
Note that it follows from these conditions that for each $U \in \F'$,
there is exactly one $i \in I$ such that $0 < \mu_i(U) < \infty$.

There is an obvious analogue of the map $\FCP$ mapping dimensionally
ordered systems to cps's.  Namely, let $\FDC$ map the dimensionally
ordered system $(W,\F,\F',\{\mu_i: i \in I\})$ to the cps
$(W,\F,\F',\mu)$, where $\mu(V \mid U) = \mu_i(V \mid U)$, where $i$ is the unique
element of $I$ such that $0 < \mu_i(U) < \infty$.  R\'{e}nyi shows that
$\FDC$ is a bijection from dimensionally ordered systems to cps's
where the set $\F'$ is closed under finite unions.  (Cs\'{a}sz\'{a}r
\citeyear{Csaszar55} extends this result to cases where the set $\F'$ is
not necessarily closed under finite unions.)
R\'{e}nyi assumes that all measures involved are countably additive and
that $\F$ is a $\sigma$-algebra, but these are inessential assumptions.
That is, his proof goes through without change if $\F$ is an algebra and
the measures are additive; all that happens is that the resulting
conditional probability measure is additive rather than
$\sigma$-additive.  

It is critical in R\'{e}nyi's framework that the $\mu_i$'s are arbitrary
measures, and not just probability measures.  His result does not hold
if the $\mu_i$'s are required to be probability measures. In the case of 
finitely additive measures, the Popper space constructed in
Example~\ref{counter1} already shows why.  It corresponds to a
dimensionally ordered space $(\mu_1,\mu_2)$ where $\mu_1(U)$ is 1 if $U$ is
cofinite and 0 if $U$ is finite and $\mu_2(U) = |U|$ (\ie
the measure of a set is its cardinality).  It cannot be captured by a
dimensionally ordered space where all the elements are probability
measures, for the same reason that it is not the image of an SLPS under
$\FCP$.  (R\'{e}nyi \citeyear{Renyi56} actually provides a
general characterization of when the $\mu_i$'s can be taken to be
(countably additive) probability measures.)
Another example is provided by the Popper space considered in
Example~\ref{counter2}.  This corresponds to the dimensionally ordered
system $\{\mu_\beta: \beta \in \IN \union \{\infty\}\}$, where 
$$
\mu_n(U) =
\left \{ \begin{array}{ll}
0 &\mbox{if $\max(U) < n$}\\
1 &\mbox{if $\max(U) = n$}\\
\infty &\mbox{if $\max(U) > n$},
\end{array} \right.
$$
where $\max(U)$ is taken to be  $\infty$ if $U$ is cofinite.

Krauss \citeyear{Kr68} restricts to Popper algebras of the form $\F
\times (\F-\{\emptyset\})$; this allows him to simplify and generalize
R\'{e}nyi's analysis.  Interestingly, he also proves a representation
theorem in the spirit of R\'{e}nyi's that involves  nonstandard
probability.

Van Fraassen \citeyear{vF76} proves a
result whose assumptions are somewhat closer to Theorem~\ref{infiso}.
Van Fraassen considers what he calls  {\em ordinal families of
probability measures}.  An ordinal family over $(W,\F)$ is a sequence of
the form $\{(W_\beta,\F_\beta,\mu_\beta): \beta < \alpha\}$ such that
\begin{itemize}
\item $\union_{\beta < \alpha} W_\beta =  W$;
\item $\F_\beta$ is an algebra over $W_\beta$; 
\item $\mu_\beta$ is a probability measure with domain $\F_\beta$;  
\item $\union_{\beta < \alpha} \F_\beta = \F$;
\item if $U \in \F$ and $V \in \F_\beta$, then $U \inter V \in \F_\beta$;
\item if $U \in \F$, $U \inter V \in \F_\beta$, and $\mu_\beta(U \inter V) >
0$, then there exists $\gamma$ such that $U \in \F_\gamma$ and
$\mu_\gamma(U) > 0$.
\end{itemize}

Given an ordinal family $\{(W_\beta,\F_\beta,\mu_\beta): \beta < \alpha\}$ over
$(W,\F)$, consider the map $\FOP$ which associates with it the cps
$(W,\F,\F',\mu)$, where $\F' = \{U \in \F:
\mu_\gamma(U) > 0 \mbox{ for some } \gamma < \alpha\}$ and $\mu(V \mid U) =
\mu_\beta(V \mid U)$, where $\beta$ is the smallest ordinal such that $U \in
\F_\beta$ and $\mu_\beta(U) > 0$.  
Van Fraassen shows that $\FOP$ is a bijection from ordinal families
over $(W,\F)$ to Popper spaces over $(W,\F)$.  Again, for van Fraassen,
countable additivity does not play a significant role.  If $\F$ is a
$\sigma$-algebra, a {\em countably additive\/} ordinal family over
$(W,\F)$ is defined just as an ordinal family, except that now
$\F_\beta$ is a $\sigma$-algebra over $W_\beta$ for all
$\beta < \alpha$, $\mu_\alpha$ is a countably additive probability
measure, and $\F$ is
the least $\sigma$-algebra containing $\union_{\beta <
\alpha} \F_\beta$ (since $\union_{\beta < \alpha} \F_\beta$ is not in
general a $\sigma$-algebra).  
The same map
$\FOP$ is also a bijection from countably additive ordinal families
to countably additive  Popper spaces.

Spohn's result, Theorem~\ref{infiso}, can be viewed as a
strengthening of van Fraassen's result in the countably additive case,
since for Theorem~\ref{infiso} all the $\F_\beta$'s are
required to be identical.  This is a nontrivial requirement.  The fact
that it cannot be met in the case that $W$ is infinite and the measures
are not countably additive is an indication of this.

It is worth seeing  how van Fraassen's approach handles the finitely
additive examples which do not correspond to SLPS's.  
The Popper space in Example~\ref{counter1} corresponds to the ordinal
family $\{(W_n,\F_n,\mu_n): n \le \omega\}$ where, for $n < \omega$,
$W_n = \{1, \ldots, n\}$, $\F_n$ consists of all subsets of $W_n$, and
$\mu_n$ is the uniform measure, while $W_\omega = \IN$, $\F_\omega$
consists of the finite and cofinite subsets of $\IN$, and $\mu_\omega(U)$ is 1
if $U$ is cofinite and 0 if $U$ is finite.  It is easy to check that
this ordinal family has the desired properties.  
The Popper space in
Example~\ref{counter2} is represented in a similar way, using the
ordinal family $\{(W_n,\F_n,\mu_n'): n \le \omega\}$, where $\mu_n'(U)$
is 1 if $n \in U$ and 0 otherwise, while $\mu_\omega' = \mu_\omega$.  I
leave it to the reader to see that this family has the desired
properties. 
The key point to observe here is the leverage obtained by
allowing each probability measure to have a different domain.

\section{Relating LPS's to NPS's}\label{LPSNPS}

In this section, I show that LPS's and NPS's are 
isomorphic in a strong sense.  
Again, I separate the results for the finite case and the infinite case.

\subsection{The finite case}
Consider an LPS of the form $(\mu_1, \mu_2,\mu_3)$.  Roughly speaking,
the corresponding NPS should be $(1 - \epsilon - \epsilon^2) \mu_1 +
\epsilon \mu_2 + \epsilon^2 \mu_3$, where $\epsilon$ is some
infinitesimal.  That means that $\mu_2$ gets infinitesimal weight
relative to $\mu_1$ and $\mu_3$ gets infinitesimal weight relative to
$\mu_2$.  But which infinitesimal $\epsilon$ should be
chosen?  Intuitively, it shouldn't matter.  No matter which
infinitesimal is chosen, the resulting NPS should be equivalent to the
original LPS.  I now make this intuition precise.

Suppose that we want to use an LPS or an NPS to compute which of two
bounded, {\em real-valued\/} random variables has higher expected value.%
The intended 
application here is decision making, where the random variables can be
thought of as the utilities corresponding to two actions; the one with
higher expected utility is preferred. 
The idea is that two measures of
uncertainty (each of which can be an LPS or an NPS) are equivalent if
the preference order they place on (real valued) random variables
(according to their expected value) is the same.
I consider only random variables with countable range.  This restriction
both makes the 
exposition simpler and avoids having to define, for example, integration
with respect to an NPS.  Note that, given an LPS $\vecmu$, the
expected value of a random variable $X$ is $\sum_x x \vecmu(X=x)$, where
$\vecmu(X=x)$ is a sequence of probability values and the multiplication
and addition are pointwise.  Thus, the expected value is a sequence;
these sequences can be compared using the lexicographic order $<_L$
defined in Section~\ref{LPSdef}.  If $\nu$ is either an LPS or NPS, 
then let $E_\nu(X)$ denote the expected value of random variable $X$
according to $\nu$.

\dfn\label{aeq} If each of $\nu_1$ and $\nu_2$ is either an NPS over
$(W,\F)$ or an 
LPS over $(W,\F)$, then $\nu_1$ is {\em equivalent to\/} $\nu_2$,
denoted $\nu_1 \aeq \nu_2$, if, for all real-valued random variables $X$
and $Y$ 
measurable with respect to $\F$,  $E_{\nu_1}(X) \le E_{\nu_1}(Y)$ iff
$E_{\nu_2}(X) \le E_{\nu_2}(Y)$.  
(If $X$ has countable range, which is the only case I consider here, then
$X$ is measurable with respect to $F$ iff $\{w: X(w) = x\} \in \F$ for
all $x$ in the range of $X$.)%
\footnote{As pointed out by Adam Brandenburger and Eddie Dekel,
this notion of equivalence is essentially the same as one
implicitly used by BBD.  They work with preference orders on
Anscombe-Aumann acts \cite{AA63}, that is, functions from states to
probability measures on prizes.  Fix a utility function $u$ on prizes.  Then
take $\nu_1 \sim_u \nu_2$ if the preference order on acts generated by
$\nu_1$ and $u$ is the same as that generated by $\nu_2$ and $u$.
It is not hard to show that this notion of equivalence is independent of
the choice of utility function; if $u$ and $u'$ are two utility
functions on prizes, then $\nu_1 \sim_u \nu_2$ iff $\nu_1 \sim_{u'}
\nu_2$.  Moreover, $\nu_1 \sim_u \nu_2$ iff $\nu_1 \aeq \nu_2$.
The advantage of the notion of equivalence used here is that it is
defined without the overhead of preference orders on acts.}
\edfn

This notion of equivalence satisfies analogues of the two key
properties of the map $\FCP$ considered at the beginning of Section~\ref{FCP}.
\pro\label{FCPaeq} 
If $\nu \in \NPS(W,\F)$, $\vecmu \in \LPS(W,\F)$, and
$\nu \aeq \vecmu$, then $\nu(U) > 0$ iff $\vecmu(U) > \vec{0}$
Moreover, if $\nu(U) > 0$, then $\stand{\nu(V \mid U)} = \mu_j(V \mid U)$, where
$\mu_j$ is the first probability measure in $\vecmu$ such that $\mu_j(U)
> 0$. \epro

As the next result shows,
for SLPS's, the $\aeq$-equivalence classes
are singletons, even if the set of worlds is infinite. (This is not true
for LPS's in general. 
For example, $(\mu,\mu) \aeq (\mu)$.)  This can be viewed as providing
more motivation for the use of SLPS's.

\pro\label{motivation} If $\vecmu, \vecmu' \in \SLPS(W,\F)$, then
$\vecmu \aeq \vecmu'$ 
iff $\vecmu = \vecmu'$.  
\epro

The next result justifies restricting to finite LPS's if the state
space is finite.  
Given an algebra $\F$, let $\Bas(\F)$ consist of the
{\em basic sets\/} in $\F$, that is, the nonempty sets $\F$ that themselves
contain no nonempty subsets in $\F$.  Clearly the sets in $\Bas(\F)$ are
disjoint, so that $|\Bas(\F)| \le |W|$.  If all sets are measurable, then
$\Bas(\F)$ consists of the singleton subsets of $W$.  If $W$ is finite,
it is easy to see that all sets in $\F$ are finite unions of the sets in
$\Bas(\F)$.   

\pro\label{finiteeq} If $W$ is finite, then every LPS over $(W,\F)$ is
equivalent to an LPS of length at most $|\Bas(\F)|$. \epro

I can now define the bijection that relates NPS's
and LPS's.  Given $(W,\F)$, let $\LPS(W,\F)/\naeq$ be the equivalence
classes of $\aeq$-equivalent LPS's over $(W,\F)$; similarly, let 
$\NPS(W,\F)/\naeq$ be the equivalence classes of $\aeq$-equivalent NPS's
over $(W,\F)$.   Note that in $\NPS(W,\F)/\naeq$, it is possible that
different nonstandard probability measures could have different ranges.
For this section, without loss of generality, I could also fix the range
of all NPS's to be 
the nonstandard model 
$\IR(\epsilon)$ discussed in Section~\ref{NPSdef}.  However, in the infinite
case, it is not possible to restrict to a single nonstandard model, so I
do not do so here either, for uniformity.

Now define the mapping $\FLN$ from $\LPS(W,\F)/\naeq$ to $\NPS(W,\F)/\naeq$
pretty much as suggested at the beginning of this subsection:
If $[\vecmu]$ is an equivalence class of LPS's, then choose a
representative $\vecmu' \in [\vecmu]$ with finite length.   
Fix an infinitesimal $\epsilon$.
Suppose that 
$\vecmu' = (\mu_0, \ldots, \mu_k)$.
Let $\FLN([\vecmu]) = [(1 - \epsilon -
\cdots - \epsilon^{k}) \mu_0 + \epsilon \mu_1 + \cdots + \epsilon^k \mu_k]$.

\thm\label{lpsnps} If $W$ is finite, then 
$\FLN$ is a bijection from $\LPS(W,\F)/\naeq$ to $\NPS(W,\F)/\naeq$
that preserves equivalence (that is, each NPS in $\FLN([\vecmu])$ is
equivalent to $\vecmu$).
\ethm

\prf It is easy to check that if $\vecmu = (\mu_0, \ldots, \mu_k)$, then
$\vecmu \aeq (1 - \epsilon -
\cdots - \epsilon^{k}) \mu_0 + \epsilon \mu_1 + \cdots + \epsilon^k
\mu_k$ (see Lemma~\ref{aeqchar} in the appendix for a formal proof).
It follows that $\FLN$ is an injection from 
$\LPS(W,\F)/\naeq$ to $\NPS(W,\F)/\naeq$.    To show that $\FLN$ is a
surjection, we must essentially construct an inverse map; that is, given
an NPS $(W,\F,\nu)$ where $W$ is finite, we must find an LPS $\vecmu$
such that $\vecmu \aeq \nu$.  The idea is to find 
a finite collection $\mu_0, \ldots, \mu_k$ of (standard)
probability measures, where $k \le |W|$, and nonnegative nonstandard reals
$\epsilon_0, \ldots, \epsilon_k$ such that
$\stand{\epsilon_{i+1}/\epsilon_i} = 0$ and $\nu = \epsilon_0 \mu_0 +
\cdots + \epsilon_k\mu_k$.  A straightforward argument then shows that 
$\nu \aeq \vecmu$ and $\FLN([\vecmu]) = [\nu]$.  I leave details to
the appendix. \eprf

BBD \citeyear{BBD1} also relate nonstandard probability measures and
LPS's under the assumption that the state space is finite,
but there are some significant technical differences between the way
they relate them and the approach taken here.
BBD prove representation theorems
essentially showing that a preference order on lotteries 
can be represented by a standard utility function on lotteries and an
LPS iff it
can be represented by a standard utility function on lotteries and an NPS.
Thus, they show that NPS's and LPS's are equiexpressive in terms of
representing preference orders on lotteries.  
The difference between 
BBD's result and Theorem~\ref{lpsnps} is essentially a matter of
quantification.  BBD's  result can be viewed as showing that, given an
LPS, for each utility function on lotteries, there is an NPS that
generates the same preference order on lotteries for that particular
utility function.  In principle, the NPS might depend on the utility
function.  More precisely, for a fixed LPS $\vecmu$, all
that follows from their result is that for each utility function $u$, there
is an NPS $\nu$ such that $(\vecmu,u)$ and $(\nu,u)$ generate the same
preference order on lotteries.   Theorem~\ref{lpsnps} says that, given
$\vecmu$, there is an NPS $\nu$ such that $(\vecmu,u)$ and $(\nu,u)$
generate the same preference on lotteries for {\em all\/} utility
functions $u$.

\subsection{The infinite case}

An LPS over an infinite state space $W$ may not be equivalent to any
finite LPS.   However, ideas analogous to those used to prove
Proposition~\ref{finiteeq} can be used to provide a bound on the length
of the minimal-length LPS's in an equivalence class.

\pro\label{infiniteeq} Every LPS over $(W,\F)$ is
equivalent to an LPS over $(W,\F)$ of length at most $|\F|$. \epro

The first step in relating LPS's to NPS's is to show that, just as in
the finite case, for every LPS $(\mu_\beta: \beta < \alpha)$ of length
$\alpha$, there is an equivalent NPS $\nu$.  The idea will
be to 
set
$\nu = (1 - \sum_{0 < \beta < \alpha} \epsilon^{n_\beta}) +
\sum_{0 < \beta < \alpha} \epsilon_{n_\beta} \mu_\beta$.  In the finite
case, we could take $n_\beta = \beta$.  This worked because 
each $\beta$ was finite, and the field $\IR(\epsilon)$ includes
$\epsilon^j$ for each integer $j$.   But now, since $\alpha$ may be
greater than $\omega$, we cannot just take $n_\beta = \beta$.
To get this idea to work in the infinite setting, I consider a 
\emph{nonstandard} model of the integers, which includes an ``integer''
corresponding to all the ordinals less than $\alpha$.  I then 
construct a field that includes $\epsilon^{n_\alpha}$ even for
these nonstandard integers $n_\alpha$.

A {\em nonstandard model of the integers\/} is a 
model that contains the integers and satisfies every property of the
integers expressible in first-order logic.
It follows easily from the compactness theorem of first-order
logic \cite{Enderton} that, given an ordinal $\alpha$, there exists a
nonstandard model 
$I^\alpha$ 
of the integers $I^\alpha$ that includes elements
$n_\beta$, $\beta < 
\alpha$, such that $n_j = j$ for $j <\omega$ and $n_\beta < n_{\beta'}$
if $\beta < \beta'$.  (Note that since $I^\alpha$ satisfies all the
properties of the integers, it follows that if $n_\beta < n_{\beta'}$,
then $n_{\beta'} - n_\beta \ge 1$, a fact that will be useful later.) 
The compactness theorem says that, given a collection of 
formulas, if each finite subset has a model, then so does the whole set.
Consider a language with a function $+$ and constant symbols for each
integer, together with constants ${\bf n}_\beta$, $\beta < \alpha$.
Consider the collection of first-order formulas in this language
consisting of all the formulas true of the integers, together with the
formulas ${\bf n}_i = i$ for $i < \omega$ and ${\bf n}_\beta < {\bf
n}_{\beta'}$, for all $\beta < \beta' < \alpha$.
Clearly any finite subset of this set has a model---namely, the
integers.  Thus, by compactness, so does the full set.  Thus, for each
ordinal $\alpha$, there is a model $I^\alpha$ 
with the required properties.


Given $\alpha$, I now construct a field $\IR(I^\alpha)$
that includes $\epsilon^n$ for each ``integer'' $n \in I^\alpha$.
To explain the construction,
it is best to first consider $\IR(\epsilon)$ in a
little more detail.  Since $\IR(\epsilon)$ is a field, once it includes
$\epsilon$, it must include $p(\epsilon)$, 
where $p$ is a polynomial with real coefficients.  To ensure the every
nonzero element of $\IR(\epsilon)$ has an inverse, we need not just
finite polynomials in $\epsilon$, but \emph{infinite} polynomials in
$\epsilon$.  The inverse of a polynomial in $\epsilon$ can then be
computer using standard ``formal'' division of polynomials.  
Moreover, the leading coefficient of the polynomial can be negative.  
Thus, the inverse of $\epsilon^3$ is, not surprisingly, $\epsilon^{-3}$; 
the inverse of $1-\epsilon$ is $1 + \epsilon + \epsilon^2 + \ldots$. 

The field $\IR(I^\alpha)$ also includes polynomials in $\epsilon$, but
now the exponents are not just integers, but elements of 
$I^\alpha$.  Since a field is closed under multiplication, if it
contains $\epsilon^{n_1}$ and $\epsilon^{n_2}$, it must also include
their product.  Since $I^\alpha$ satisfies all the properties of the
integers, if it includes $n_1$ and $n_2$, it also includes an element
$n_1 + n_2$, and we can take $\epsilon^{n_1} \times \epsilon^{n_2} =
\epsilon^{n_1 + n_2}$. Formally, let $\IR(I^\alpha)$ be the 
non-Archimedean model defined as follows:
$\IR(I^\alpha)$ consists of all polynomials of the form
$\sum_{n \in J} r_n \epsilon^{n}$, where $r_n$ is a standard real,
$\epsilon$ is an infinitesimal, and $J$ is a \emph{well-founded} subset
of $I^\alpha$.  (Recall that a set is well founded if it has no
infinite descending sequence; thus, the set of integers is not well
founded, since $\ldots -3 < -2 < -1$ is an infinite descending
sequence.  The reason I require well foundedness will be clear shortly.)
We can identify the standard real $r$ with the polynomial 
$r \epsilon^0$.  

The polynomials in $\IR(I^\alpha)$ can be added and
multiplied using the standard rules for addition and multiplication of
polynomials. 
It is easy to check that
the result of adding or multiplying two
polynomials is another polynomial in $\IR(I^\alpha)$.  In particular, if
$p_1$ and $p_2$ are 
two polynomials, $N_1$ is the set of exponents of $p_1$, and 
$N_2$ is the set of exponents of $p_2$, then the
exponents of $p_1 + p_2$ lie in $N_1 \union N_2$, while the
exponents of $p_1p_2$ lie in the set $N_3 = \{n_1 + n_2: 
n_1 \in N_1, n_2 \in N_2\}$.  Both $N_1 \union N_2$ and $N_3$ are easily 
seen to be well founded if $N_1$ and $N_2$ are.  Moreover, for each
expression $n_1 + n_2 \in N_3$, it 
follows from the well-foundedness of $N_1$ and $N_2$ that there are only
finitely many pairs $(n,n') \in N_1 \times N_2$ such that $n+n' = n_1 +
n_2$, 
so the coefficient of $\epsilon^{n_1 + n_2}$ in $p_1p_2$ is well defined.
Finally, each polynomial (other than 0) has an
inverse that can be computed using standard ``formal'' division of
polynomials;  I leave the details to the reader.  
This step is where the well foundedness comes in.  The formal division
process cannot be applied to a polynomial with coefficients that are not
well founded, such as $\cdots + \epsilon^{-3} + \epsilon^{-2} +
\epsilon^{-1}$.  An element 
of $\IR(I^\alpha)$ is {\em positive\/} if its leading coefficient is
positive.  Define an order $\le$ on  $\IR(I^\alpha)$ by taking $a \le b$ if 
$b-a$ is positive. 
With these definitions, $\IR(I^\alpha)$ is a non-Archimedean field.  

Given $(W,\F)$, let $\alpha$ be 
the minimal ordinal whose cardinality is greater than 
or equal to 
$|\F|$. 
By construction, $I^\alpha$ has elements $n_\beta$ for all $\beta <
\alpha$ such that 
$n_i = i$ for $i < \omega$ and $n_\beta < n_{\beta'}$ if $\beta <
\beta' < \alpha$.  
I now define a map $\FLN$ from
$\LPS(W,\F)/\naeq$ to $\NPS(W,\F)/\naeq$ just as suggested earlier.
In more detail, given an equivalence class $[\vecmu] \in \LPS(W,\F)$, by
Proposition~\ref{infiniteeq}, there exists $\vecmu' \in
[\vecmu]$ such that $\vecmu'$ has length $\alpha' \le \alpha$.  
Let $\nu = (1 - \sum_{0 < \beta < \alpha} \epsilon^{n_\beta})\mu_0 +
\sum_{0 < \beta < \alpha} \epsilon_{n_\beta} \mu_\beta'$.
By definition, $\sum_{0 < \beta < \alpha} \epsilon^{n_\beta} \in
\IR(I^\alpha)$  (the set
of exponents is well ordered since the ordinals are well ordered), hence
so is $(1 - \sum_{0 < \beta < \alpha} \epsilon^{n_\beta})$.
The elements $\epsilon^{n_\beta}$ for $\beta \le \alpha$ are also all
in $\IR(I^\alpha)$.  It easily follows that $\nu$ is nonstandard
probability measure over the field $\IR(I^\alpha)$.  As observed
earlier, if $\beta' < \beta$, then $\beta - \beta' \ge 1$, so
$\epsilon^{n_\beta'}$ is infinitesimally smaller than
$\epsilon^{n_\beta}$.  Arguments essentially identical to those of
Lemma~\ref{aeqchar} in the appendix can be
used to show that $\nu \aeq \vecmu'$.  
Define $\FLN[\vecmu] = [\nu]$.
The following result is immediate.

\thm\label{injection} $\FLN$ is an injection from $\LPS(W,\F)/\naeq$ to
$\NPS(W,\F)/\naeq$ that preserves equivalence.  \ethm

What about the converse?  Is it the case that for every NPS there is an
equivalent LPS?  
The technique for finding an equivalent LPS used in the finite case
fails.  There is no obvious way to find a well-ordered sequence
of standard probability measures $\mu_0, \mu_1, \ldots$ and a sequence
of nonnegative nonstandard reals $\epsilon_0, \epsilon_1, \ldots$ such
that $\stand{\epsilon_{\beta+1}/\epsilon_\beta} = 0$ and 
$\nu = \epsilon_0 \mu_0 + \epsilon_1 \mu_1 + \cdots$.  As the following
example shows, this is not an accident.  
There exists NPSs that are not equivalent to any LPS. 

\xam\label{counter3} As in Example~\ref{counter1}, let $W = \IN$, the
natural numbers, let $\F$ consist of the finite and cofinite subsets of
$\IN$,
and let $\F' = \F - \{\emptyset\}$.  Let $\nu^1$ be an NPS with
range $\IR(\epsilon)$, where $\nu^1(U) = |U|\epsilon$ if $U$ is finite and
$\nu^1(U) = 1 - |\overline{U}|\epsilon$ if $U$ is cofinite
(as usual, $\overline{U}$ denotes the complement of $U$, which in this
case is finite).
This is
clearly an NPS, and it corresponds to the cps $\mu^1$ of
Example~\ref{counter1}, in the sense that $\stand{\nu^1(V \mid U)}
 = \mu^1(V \mid U)$ for all $V \in \F$, $U\in \F'$.  Just as in
Example~\ref{counter1}, it can be shown that there is 
no LPS $\vecmu$ such that $\nu^1 \aeq \vecmu$.

To see the potential relevance of this setup, suppose that 
a natural number is chosen at random and,
intuitively, all numbers are equally likely to be chosen.  An agent may
place a bet 
on the number being in a finite or cofinite set.  Intuitively, the agent
should prefer a bet on a set with larger cardinality.  More precisely,
if $U_1$ and $U_2$ are two sets in the algebra, the agent should prefer
a bet on $U_1$ over a bet on $U_2$ iff
(a) $U_1$ and $U_2$ are both cofinite and the complement of $U_1$ has
smaller cardinality than that of $U_2$, (b) $U_1$ is cofinite and $U_2$
is finite, or (c) $U_1$ and $U_2$ are both finite, and $U_1$ has larger
cardinality than $U_2$.  These preferences on acts or bets
should translate to statements of likelihood.  
The NPS captures these preferences directly; they cannot
be captured in an LPS.  The cps of Example~\ref{counter1} captures (b)
directly, and (c) indirectly: when conditioning on any finite set that
contains $U_1 \union U_2$, the probability of $U_1$ will be higher than
that of $U_2$.  
\exam

\subsection{Countably additive nonstandard probability
measures}\label{countableadditivity} 

Do things get any better if countable additivity is required?
To answer this question, I must first make precise what countable
additivity means in the context of non-Archimedean fields. 
To understand the issue here, recall that for the standard real numbers, 
every bounded nondecreasing sequence has a unique least upper bound, which
can be taken to be its limit.  Given a countable sum each of whose terms
is nonnegative, the partial sums form a nondecreasing sequence.
If the partial sums are bounded (which they are if the terms in the sums
represent the probabilities of a pairwise 
disjoint collection of sets), then the limit is well defined.

None of the above is true in the case of non-Archimedean fields.  For a
trivial counterexample,  
consider the sequence $\epsilon, 2 \epsilon, 3 \epsilon, \ldots$.
Clearly this sequence is bounded (by any positive real number), but it
does not have a least upper bound.  For a more subtle example, consider
the sequence $1/2, 3/4, 7/8, \ldots$ in the field $\IR(\epsilon)$.  Should
its limit be 1?  While this does not seem to be an unreasonable choice,
note that 1 is not the least upper bound of the sequence.  For example,
$1-\epsilon$ is greater than every term in the sequence, and is less
than 1.  So are $1-3\epsilon$ and $1 - \epsilon^2$.  Indeed, this
sequence has no least upper bound in $\IR(\epsilon)$.

Despite these concerns, I define limits in 
$\IR(I^*)$ pointwise.  That is, 
a sequence $a_1, a_2, a_3, \ldots$ in $\IR(I^*)$
converges to $b \in \IR(I^*)$ if, for every $n \in I^*$, the
coefficients of $\epsilon^n$ in $a_1, a_2, a_3, \ldots$ converge to the
coefficient of $\epsilon^n$ in $b$. (Since the coefficients are standard
reals, the notion of convergence for the 
coefficients is just the standard definition of convergence in the reals.
Of course, if $\epsilon^n$ does not appear explicitly, its coefficient
is taken to be 0.) 
Note that here and elsewhere I use the letters $a$ and $b$ (possibly with
subscripts)  to denote (standard) reals, and $\epsilon$ to denote an
infinitesimal.  
As usual, $\sum_{i=1}^\infinity a_i$ is taken to be $b$ if 
the sequence of partial sums $\sum_{i=1}^n a_i$ converges to $b$.
Note that, with this notion of convergence, $1/2, 3/4, 7/8, \ldots$
converges to 1 even though 1 is not the least upper bound of the
sequence.%
\footnote{For those used to thinking of convergence in topological
terms, what is going on here is that the topology corresponding to this
notion of convergence is not Hausdorff.}  
I discuss the consequences of this choice further in
Section~\ref{discussion}.

With this notion of countable sum, it makes perfect sense to consider
countably-additive nonstandard probability measures.  If $\F$ is a
$\sigma$-algebra and $\LPS^c(W,\F)$ and $\NPS^c(W,\F)$ denote the
countably additive LPS's and NPS's on $(W,\F)$, respectively, then
Theorem~\ref{injection} can be applied with no change in proof to
show the following.

\thm\label{injection1}  $\FLN$ is an injection from $\LPS^c(W,\F)/\naeq$
to $\NPS^c(W,\F)/\naeq$.
\ethm

However, as the following example shows, even with the requirement of
countable additivity, there are nonstandard probability measures that
are not equivalent to any LPS.

\xam\label{counter4} Let $W = \{w_1, w_2, w_3, \ldots\}$, and let $\F =
2^W$.  Choose any nonstandard $I^*$ and fix an infinitesimal $\epsilon$
in $\IR(I^*)$.
Define an NPS $(W,\F,\nu)$ with range $\IR(I^*)$
by taking $\nu(w_j) = a_j + b_j \epsilon$, where $a_j = 1/2^j$, $b_{2j-1} =
1/2^{j-1}$, and $b_{2j} = -1/2^{j-1}$, for $j = 1, 2, 3, \ldots$.
Thus, the probabilities of $w_1, w_2, \ldots$ are characterized by the
sequence $1/2 + \epsilon, 1/4 - \epsilon, 1/8 + \epsilon/2, 1/16 -
\epsilon/2, 1/32 + \epsilon/4, \ldots$.  For $U \subseteq W$, define
$\nu(U) = \sum_{\{j: w_j \in U\}} a_j + \epsilon \sum_{\{j: w_j \in U\}}
b_j$.  It is easy to see that these sums are well-defined.  
These likelihoods correspond to preferences.  For example, an agent
should prefer a bet that gives a payoff of 1 if $w_2$ occurs and 0 otherwise
to a bet that gives a payoff of 4 if $w_4$ occurs and 0 otherwise.
As I show in the appendix (see Proposition~\ref{counter}), there
is no LPS $\vecmu$ over $(W,\F)$ such that $\nu \aeq
\vecmu$. 
\exam

Roughly speaking, the reason that $\nu$ is not equivalent to
any LPS in Example~\ref{counter4} is that the ratio between $a_j$ and
$b_j$ in the definition of $\nu$ (i.e., the ratio 
between the
``standard part'' of
$\nu(w_j)$ and the ``infinitesimal part'' of $\nu(w_j)$) 
goes to zero.  This can be generalized so as to give 
a condition on nonstandard probability measures that 
is necessary and sufficient to guarantee that they can be represented by
an LPS.
However, the condition is rather technical and I have not found an
interesting interpretation of it, so I do not pursue it here.

\section{Relating Popper Spaces to NPS's}\label{PopperNPS}
Consider the map $\FNP$ from nonstandard probability spaces to Popper
spaces such that $\FNP(W,\F,\nu) = (W,\F,\F',\mu)$, where 
$\F' = \{U: \nu(U) \ne 0\}$ and $\mu(V \mid U) = \stand{\nu(V \mid U)}$ for $V \in
\F$, $U \in \F'$.  I leave it to the reader to check that
$(W,\F,\F',\mu)$ is indeed a Popper space.  
This is arguably the most natural map; for example, it is easy to check
that $\FNP \circ \FSN = \FCP$, where $\FSN$ is the restriction of $\FLN$
to SLPSs.  (Note that $\FLN$ is well-defined on SLPS's, since if
$\vecmu$ is an SLPS, by Proposition~\ref{motivation}, $[\vecmu] =
\{\vecmu\}$.)  

We might hope that $\FNP$ is a bijection from $\NPS(W,\F)/\naeq$ to
$\Popper(W,\F)$.  As I show shortly, it is not.  To understand $\FLN$
better, define an equivalence relation $\simeq$ on $\NPS(W,\F)$ (and
$\NPS^c(W,\F)$) by taking $\nu_1 \simeq \nu_2$ if $\{U: \nu_1(U) = 0\} =
\{U: \nu_2(U) = 0\}$ and $\stand{\nu_1(V \mid U)} = \stand{\nu_2(V \mid
U)}$ for 
all $V, U$ such that $\nu_1(U) \ne 0$.  
Thus, $\simeq$ essentially says that infinitesimal differences between
conditional probabilities do not count.
Let $\NPS/\!\simeq$
(\respc $\NPS^c/\!\simeq$) consist of the
$\simeq$ equivalence classes in $\NPS$ (\respc $\NPS^c$).  Clearly
$\FNP$ is well defined as a map from $\NPS/\!\simeq$ to $\Popper(W,\F)$
and 
from $\NPS^c/\!\simeq$ to $\Popper^c(W,\F)$.  As the following result
shows, $\FNP$ is actually a bijection from $\NPS^c/\!\simeq$ to
$\Popper^c(W,\F)$.

\thm\label{FNP} $\FNP$ is a bijection from $\NPS(W,\F)/\!\simeq$ to
$\Popper(W,\F)$ and from $\NPS^c(W,\F)/\!\simeq$ to $\Popper^c(W,\F)$.
\ethm

\prf 
It is easy to see that $\FNP$ is an injection.
In the countable case, the inverse map can be defined using earlier results.
If $(W,\F,\F',\mu) \in \Popper^c(W,\F)$,  by
Theorem~\ref{infiso}, 
there is a countably additive SLPS $\vec{\mu}'$ such that
$\FCP((W,\F,\vec{\mu}')) = (W, \F,\F', \mu)$.  By
Theorem~\ref{injection}, there is some 
$(W,\F,\nu) \in \NPS^c(W,\F)$ such that $\nu \aeq \vecmu'$.  It is not
hard to show that $\FNP(W,\F,\nu) = (W,\F,\F',\mu)$; see the appendix
for details.  Showing that $\FNP$ is a surjection in the finitely
additive case requires more work; again, see the appendix for details.
\eprf

McGee \citeyear{McGee94} proves essentially the same result as
Theorem~\ref{FNP} in the case that $\F$ is an algebra (and the measures
involved are not necessarily countably additive).  McGee
\citeyear[p.~181]{McGee94} says that his 
result shows that ``these two approaches amount to the same thing''.
However, this is far from clear.   The $\simeq$ relation is rather
coarse.  In particular, it is coarser than~$\aeq$.

\pro\label{simeqvsaeq} If $\nu_1 \aeq \nu_2$ then $\nu_1 \simeq \nu_2$. 
\epro

The converse of Proposition~\ref{simeqvsaeq} does not hold in general.
As a result,
the $\simeq$ relation identifies nonstandard measures that behave quite
differently in decision contexts.
This difference already arises in finite spaces, as the following example
shows.

\xam\label{McGee}
Suppose $W =
\{w_1,w_2\}$.  Consider the nonstandard probability measure $\nu_1$ such that
$\nu_1(w_1) = 1/2 + \epsilon$ and $\nu_1(w_2) = 1/2 - \epsilon$.  (This is
equivalent to the LPS $(\mu_1,\mu_2)$ where $\mu_1(w_1) = \mu_2(w_2) =
1/2$, $\mu_2(w_1) = 1$, and $\mu_2(w_2) = 0$.)  
Let $\nu_2$ be the nonstandard probability measure such that $\nu_2(w_1)
= \nu_2(w_2) = 1/2$.  Clearly $\nu_1 \simeq \nu_2$.  However, it is not
the case that $\nu_1 \aeq \nu_2$.  
Consider the two
random variables $\chi_{\{w_1\}}$ and $\chi_{\{w_2\}}$.
(I use the notation $\chi_U$ to denote the indicator function for $U$; 
that is, $\chi_U(w) = 1$ if $w \in U$ and $\chi_U(w) = 0$ otherwise.)
According to $\nu_1$, the 
expected value of $\chi_{\{w_1\}}$ is (very slightly) higher than that of
$\chi_{\{w_2\}}$. 
According to $\nu_2$, $\chi_{\{w_1\}}$ and $\chi_{\{w_2\}}$ have the
same expected value.  Thus, $\nu_1 \not\aeq \nu_2$.
Moreover, it is easy to see that there 
is no Popper measure $\mu$ on $\{w_1,w_2\}$ that can make the same
distinctions with respect to $\chi_{\{w_1\}}$ and $\chi_{\{w_2\}}$ as
$\nu_1$, no matter how we define expected value with respect to a
Popper measure.   According to $\nu_1$, although the expected value of
$\chi_{\{w_1\}}$ is higher than that of $\chi_{\{w_2\}}$, the expected
value of $\chi_{\{w_1\}}$ is less than 
that of $\alpha \chi_{\{w_2\}}$ for any (standard) real $\alpha > 1$.
There is no Popper measure with this behavior.
\exam

More generally, in finite spaces, Theorem~\ref{FCPfin} shows that 
Popper spaces are equivalent to SLPS's, while
Theorem~\ref{lpsnps} shows that $\LPS(W,\F)/\naeq$ is equivalent to 
$\NPS(W,\F)/\naeq$.  By Proposition~\ref{motivation},
$\SLPS(W,\F)/\naeq$ is essentially identical to $\SLPS(W,\F)$ (all the
equivalence classes in $\SLPS(W,\F)/\naeq$ are singletons), 
so in finite spaces, the gap in expressive power between Popper spaces
and NPS's essentially amounts to the gap between $\SLPS(W,\F)$ and
$\LPS(W,\F)/\naeq$.  This gap is nontrivial.  For example, there is no
SLPS equivalent to the LPS $(\mu_1,\mu_2)$ that represents the NPS in
Example~\ref{McGee}. 


\section{Independence}\label{sec:indep}
The notion of independence is fundamental.  As I show in this section, the
results of the previous sections sheds light on various notions of
independence considered in the literature for LPS's and (variants of)
cps's.  I first consider independence for events and then independence
for random variables.  I then relate my definitions to those of BBD,
Hammond, and Kohlberg and Reny \citeyear{KR97}.

Intuitively, event $U$ is independent of $V$ if learning $U$ gives no
information about $V$.  Certainly if learning $U$ gives no information
about $V$, then if $\mu$ is an arbitrary probability measure, we would
expect that $\mu(V \mid U) = \mu(V)$.  Indeed, this is often taken as the
definition of $V$ being independent of $U$ with respect to $\mu$.  
If standard probability measures are used, conditioning is not
defined if $\mu(U) = 0$.  In this case, $U$ is still considered
independent of $V$.  As is well known, if $U$ is independent of $V$,
then $\mu(U \inter V) = \mu(V) \times \mu(U)$ and $V$ is independent of
$U$, that is, $\mu(U \mid V) = \mu(U)$.  Thus, independence of events with
respect to a probability measure can be
defined in any of three equivalent ways.  Unfortunately, these
definitions are not equivalent for other representations of uncertainty
(see \cite[Chapter 4]{Hal31} for a general discussion of this issue).

The situation is perhaps simplest for nonstandard probability measures.%
\footnote{Although I talk about $U$ being independent of $V$ with
respect to a nonstandard measure $\nu$, technically I should talk about
$U$ being independent of $V$ with respect to an NPS $(W,\F,\nu)$, for 
$U, V \in \F$.  I continue to be sloppy at times, reverting to more
careful notation when necessary.}
In this case, the three notions coincide, for exactly the same reasons
as they do for standard probability measures.  However, independence is
perhaps too strong a notion in some ways.  In particular, nonstandard
measures that are equivalent do not in general agree on independence, as
the following example shows.
\xam\label{xam:approximatelyindep}
Suppose that $W = \{w_1, w_2, w_3, w_4\}$.  Let
$\nu_i(w_1 ) = 1 - 2 \epsilon + \epsilon_i$, $\nu_i(w_2) = \nu_i(w_3) =
\epsilon - \epsilon_i$, and $\nu_i(w_4) = \epsilon_i$,  for $i = 1,
2$, where $\epsilon_1 = \epsilon^2$ and $\epsilon_2 = \epsilon^3$.  If
$U = \{w_2, w_4\}$ and $V = \{w_3, w_4\}$, then $\nu_i(U) = 
\nu_i(V) = \epsilon$ and $\nu_i(U \inter V) = \epsilon_i$.  It 
follows $U$ and $V$ are independent with
respect to $\nu_1$, but not with respect to $\nu_2$.   However, it is
easy to check that $\nu_1 \aeq \nu_2$.  
\exam

Example~\ref{xam:approximatelyindep} shows that independence of events in the
context of nonstandard 
measures is very sensitive to the choice of $\epsilon$, even if this
choice does not affect decision making at all.  This suggests the
following definition: $U$ is {\em approximately independent\/} of $V$ with
respect to $\nu$ if $\nu(U) \ne 0$ implies that
$\nu(V \mid U) - \nu(V)$ is infinitesimal, that is, if
$\stand{\nu(V \mid U)} = \stand{\nu(V)}$.  
Note that $U$ can be approximately independent of $V$ without
$V$ being approximately independent of $U$.  For example, consider the
nonstandard probability measure $\nu_1$ from
Example~\ref{xam:approximatelyindep}.  Let 
$V' = \{w_4\}$; 
as before, let $U = \{w_2, w_4\}$.  It is easy to check that
$\stand{\nu_1(V' \mid U)} = \stand{\nu_1(V')} = 0$, but
$\stand{\nu_1(U \mid V')} = 1$, while $\stand{\nu_1(U)} = 0$.  Thus, 
$U$ is approximately independent of $V'$ with respect to $\nu_1$, but
$V'$ is not 
approximately independent of $U$.   Similarly, $U$ can be approximately
independent of $V$ without $\overline{U}$ being approximately
independent of $V$.  For example, it is easy to check that 
$\overline{V}'$ is approximately independent of $U$ with respect to
$\nu_1$, although $V'$ is not.

A straightforward argument shows that $U$
is approximately independent of $V$ with respect to $\nu$ iff 
$\nu(U) \ne 0$ implies $\stand{(\nu(V
\inter U) - \nu(V) \times \nu(U))/ \nu(U)} = 0$, while $V$ is
approximately independent of 
$U$ with respect to $\nu$ iff the same statement holds with the roles of
$V$ and $U$ reversed.
Note for future reference that each of these requirements
is stronger than just 
requiring that $\stand{\nu(V \inter U) - \nu(V) \times \nu(U)} = 0$. 
The latter requirement is automatically met, for example, if the
probability of either $U$ or $V$ is infinitesimal.

The definition of (approximate) independence extends in a straightforward
way to (approximate) conditional independence.  $U$ is 
conditionally independent of $V$ given $V'$ with respect to a (standard
or nonstandard) probability measure $\nu$ if 
$\nu(U \inter V') \ne 0$ implies $\nu(V \mid U \inter V') = \nu(V \mid V')$.  
Again, for probability, $U$ is
conditionally independent of $V$ given $V'$ iff $V$ is conditionally
independent of $U$ given $V'$ iff $\nu(V \inter U \mid V') = \nu(V
\mid V') \times \nu(U \mid V')$.  
$U$ is approximately
conditionally independent of $V$ given $V'$ with respect to $\nu$ if
$\stand{\nu(V \mid U \inter V')} = \stand{\nu(V
\mid V')}$.   If $V'$ is taken to be $W$, the whole space, then (approximate)
conditional independence reduces to (approximate) independence.

The following proposition shows that, although independence is not
preserved by equivalence, approximate independence is.

\pro\label{indaeq} If $U$ is approximately conditionally independent of
$V$ given 
$V'$ with respect to $\nu$, and $\nu \aeq \nu'$, then
$U$ is approximately conditionally independent of $V$ given
$V'$ with respect to $\nu'$.
\epro

\prf Suppose that $\nu \aeq \nu'$.  I claim that for all events $U_1$
and $U_2$ such that $\nu_1(U_2) \ne 0$, $\stand{\nu(U_1)/\nu(U_2)} =
\stand{\nu'(U_1)/\nu'(U_2)}$.  For suppose that
$\stand{\nu(U_1)/\nu(U_2)} = \alpha$.  Then it easily follows that
$E_\nu(\chi_{U_1}) < E_\nu(\alpha'\chi_{U_2})$ for all $\alpha' > \alpha$,
and $E_\nu(\chi_{U_1}) > E_\nu(\alpha''\chi_{U_2})$ for all $\alpha'' <
\alpha$.  Thus, the same must be true for $E_{\nu'}$, and hence
$\stand{\nu'(U_1)/\nu'(U_2)} = \alpha$.   It thus follows
that $\stand{\nu (V \mid  U \inter V')} = \stand{\nu' (V \mid  U \inter
V')}$ and $\stand{\nu(V \mid V')} = \stand{\nu'(V \mid V')}$, from which
the result is immediate. \eprf

\commentout{
There is also an interesting connection between approximate independence
and independence, which will prove useful in understanding issues
involving independence between random variables.

\pro\label{indaeq1} There exists a measure $\nu'$ such
that $\nu \aeq \nu'$ and $U$ is conditionally independent of $V$ given
$V'$ with respect to $\nu'$ iff (a) both $U$ and $\overline{U}$ are
approximately conditionally independent of $V$ given $V'$ with respect
to $\nu$ and (b) both $V$
and $\overline{V}$ are approximately conditionally independent of $U$
given $V'$ with respect to $\nu$.
\epro

\prf First suppose that $\nu \aeq \nu'$ and  and that $U$ is
conditionally independent of $U$ given $V$ with respect to $\nu'$.
Then, by standard properties of independence, both $U$ and
$\overline{U}$ are conditionally independent of $V$ given $V'$ 
and both $V$ and $\overline{V}$ are conditionally independent of $U$
given $V'$.  Since conditional independence certainly implies
conditional approximate independence, the forward implication follows
from Proposition~\ref{indaeq}.

For the reverse implication, suppose that (a) both $U$ and
$\overline{U}$ are approximately conditionally independent of $V$ given
$V'$ with respect to $\nu$ and (b) both $V$
and $\overline{V}$ are approximately conditionally independent of $U$
given $V'$ with respect to $\nu$.  Suppose that $\stand{U \mid V'} =
r_1$, $\stand{V \mid V'} = r_2$
We now need to consider a number of
cases.  First, suppose that both $0 < r_1, r_2 < 1$.  
}

There is an obvious definition of independence for events for Popper spaces:
$U$ is independent of $V$ given $V'$ with respect to the Popper space
$(W,\F,\F',\mu)$ if $U \inter V' \in\F'$ implies that $\mu(V \mid U
\inter V') = \mu(V \mid V')$; if $U \inter V' \notin \F'$, then $U$ is also
taken to be independent of $V$ given $V'$.  If
$U$ is independent of $V$ given $V'$ and $V' \in \F'$, then $\mu(U
\inter V \mid V') = \mu(U \mid V') \times \mu(V \mid V')$.  However, the
converse does not necessarily hold.  Nor is it the case that if $U$ is
independent of $V$ given $V'$ then $V$ is independent of $U$ given
$V'$.  A counterexample can be obtained by taking the Popper space
arising from the NPS in Example~\ref{xam:approximatelyindep}.  Consider the
Popper space $(W,2^W,\F',\mu)$ corresponding to the NPS $(W,2^W,\nu_1)$
via the bijection $\FNP$.  It is easy to check that $U$ is independent
of $V'$ but $V'$ is not independent of $U$ with respect to this Popper
space, although $\mu(V' \inter U) = \mu(U \mid V') \times \mu(V') \ (=
0)$.  This observation is an instance of the following more general
result, which is almost immediate from the definitions:

\pro\label{pro:approximatelyindep} $U$ is approximately independent of
$V$ given $V'$ 
with respect to the NPS $(W,\F,\nu)$ iff $U$ is independent of
$V$ given $V'$ with respect to the Popper space $\FNP(W,\F,\nu)$.
\epro

How should independence be defined in LPS's?  
Interestingly, neither BBD nor Hammond define independence 
directly
for LPS's.
\commentout{
BBD \citeyear{BBD1} give three definitions of independence: two of them
are given in terms of NPS's; the third is an indirect definition in
terms of preference orders.  Hammond also works in NPS's.  Note that 
requiring that $\vecmu(V
\mid U) = \vecmu(V)$ will not work since $\vecmu \mid
U$ and $\vecmu$ are, in general, LPS's of different lengths.  Nor is
there any obvious way to define multiplication of two LPS's.  
One way 
to define independence in LPS's is to
essentially reduce the definition to that for Popper spaces.  That is,
$U$ is independent of $V$ given $V'$ with respect to the LPS
$(W,\F,\vecmu)$ if the leftmost number in the sequence $\vecmu(V \mid U
\inter V')$ is the same as the leftmost number in $\vecmu(V \mid V')$;
as usual, independence is taken to hold trivially if $\vecmu(U \inter
V') = \vec{0}$.  The following result is almost immediate from the
definitions.  

\pro\label{pro:approximatelyindep1} $U$ is independent of $V$ given $V'$
with respect to the LPS $\vecmu$ iff $U$ is approximately independent of
$V$ given $V'$ with respect to each NPS in the equivalence class
$\FLN([\vecmu])$.
\epro

\noindent Propositions~\ref{pro:approximatelyindep}
and~\ref{pro:approximatelyindep1} emphasize 
the naturalness of approximate independence in this context.
}
However, they do give definitions in terms of NPS's that can be
applied to equivalent LPS's; indeed, BBD \citeyear{BBD2} do just this
(see the discussion of BBD strong independence below).

I now consider independence for random variables.  If $X$ is a random
variable on $W$, let $\V(X)$ denote
range (set of possible values) of random variable $X$; that is, $\V(X) =
\{X(w): w \in W\}$.
Recall that I am assuming that all random variables have countable range.
Random variable $X$ is 
independent of $Y$ with respect to a standard probability measure $\mu$
if the event $X=x$ is independent of the 
event $Y=y$ with respect to $\mu$, for all $x \in \V(X)$ and $y \in \V(Y)$.  
By analogy, for nonstandard probability measures, following Kohlberg and
Reny \citeyear{KR97}, 
define $X$ and $Y$ to
be {\em weakly independent\/} with respect to $\nu$ if  $X=x$ is
approximately independent of $Y=y$ and $Y=y$ is approximately
independent of $X=x$ with respect to $\nu$ for all $x\in 
\V(X)$ and $y \in \V(Y)$.%
\footnote{Kohlberg and Reny's definition of weak independence also
requires that the joint 
range of $X$ and $Y$ be the product of the individual ranges.  That is,
for $X$ and $Y$ to be weakly independent, it must be the case that for
all $x \in \V(X)$ and $y \in \V(Y)$, there exists some $w \in W$ such
that $X(w) = x$ and $Y(w) = y$. 
Of course, this requirement could also be added to the definition 
I am proposing here; adding it would not affect any of the results
of this paper.}

For standard probability measures, it easily follows
that if $X$ is independent of $Y$, then $X \in U_1$ is independent of $Y
\in V_1$ conditional on $Y \in V_2$ and $Y \in V_1$ is independent of $X
\in U_1$ conditional on $X \in U_2$, for all $U_1, U_2 \subseteq \V(X)$
and $V_1, V_2 \subseteq \V(Y)$.   The same arguments show that this is
also true for for nonstandard probability measures.  However, the
argument breaks down for approximate independence.   

\xam\label{xam:needapproximate} Suppose that $W = \{1,2,3\} \times
\{1,2\}$. Let $X$ and $Y$ be the random variables that project onto the
first and second components of a world, respectively, so that $X(i,j) =
i$ and $Y(i,j) = j$.  Let $\nu$ be the nonstandard probability measure
on $W$ given by the following table:

\begin{center}
\begin{tabular}{| c | c | c | c |}
\hline
& $Y=1$ & $Y=2$ \\
\hline
$X=1$ & $1 - 3 \epsilon - 3\epsilon^2$ & $\epsilon$\\
\hline
$X=2$ & $\epsilon$ & $\epsilon^2$\\
\hline
$X=3$ & $\epsilon$ & $2\epsilon^2$\\
\hline
\end{tabular}
\end{center}
It is easy to check that 
$X$ and $Y$ are weakly independent
with respect to $\nu$, for all $i \in
\{1,2,3\}$, $j \in \{2,3\}$.  However, $\stand{\nu(X = 2 \mid X \in
\{2,3\} \inter Y=2)} = 1/3$, while $\stand{\nu(X=2 \mid X \in \{2,3\})}
= 1/2$.  
\exam

In light of this example, I define $X$ to be {\em approximately independent of
$\{Y_1, \ldots, Y_n\}$ with respect to $\nu$\/} if $X \in U_1$ is
approximately independent of $(Y_1 \in V_1) \inter \ldots \inter (Y_n
\in V_n)$ conditional on $(Y_1 \in V_1') \inter \ldots \inter (Y_n \in
V_n')$ with 
respect to $\nu$ for all
$U_1 \subseteq \V(X)$, $V_i, V_i' \subseteq \V(Y_i)$, and $i = 1, \ldots,
n$.   $X_1, \ldots, X_n$ are {\em approximately independent with respect
to $\nu$\/} if $X_i$ is approximately independent of $\{X_1, \ldots,
X_n\} - \{X_i\}$ with respect to $\nu$ for $i = 1, \ldots, n$.  I leave
to the reader the obvious extensions to 
conditional independence and the 
analogues of this definition for Popper spaces and LPS's.



As I said, BBD consider three notions of independence for random variables.
One is a decision-theoretic notion of stochastic independence on preference
relations on acts over $W$.  Under appropriate assumptions, it can be
shown that a preference relation is stochastically independent 
iff it can be 
represented by some (real-valued) utility function $u$ and a nonstandard
probability measure $\nu$ such that $X_1, \ldots, X_n$ are approximately
independent with respect to $\nu$ \cite{BV96}.
A second notion they consider is a weak notion of
product measure that requires only that there exist measures $\nu_1,
\ldots, \nu_n$ such that $\stand{(\nu(w_1, \ldots, w_n)} =
\stand{\nu_1(w_1) \times \cdots \nu(w_n)}$.  As we have already
observed, this notion of independence is rather weak.  Indeed, an
example in BBD shows that it misses out on some interesting
decision-theoretic behavior.  

\commentout{
Approximate independence and strong independence differ in the order of
universal and existential quantification.  $X$ and $Y$ are
approximately independent with respect to $\nu$ if, for all values $x$
and $y$ in the range of $X$ and $Y$, respectively, there is an NPS
$\nu_{xy}$ such that $\nu_{xy} \aeq \nu$ and $X=x$ and $Y=y$ are
independent with respect to $\nu_{xy}$.  On the other hand, $X$ and $Y$
are strongly independent if there exists an NPS $\nu'$ such that $\nu'
\aeq \nu$ and for all $x$ and $y$ in the range of $X$ and $Y$,
respectively, $X = x$ is independent of $Y=y$.  Clearly KR-strong
independence implies approximate independence.  As the following
example (due to Kohlberg and Reny \citeyear{KR97}) shows, in general, it 
is strictly stronger.

\xam Suppose that $W = \{1,2,3\} \times \{1,2,3\}$.
Let $X$ and $Y$ be the random variables that project onto the first and second
components of a world, respectively, so that $X(i,j) = i$ and $Y(i,j) =
j$.  Let $\nu$ be the nonstandard probability measure on $W$ given by the 
following table:

\begin{center}
\begin{tabular}{| c | c | c | c |}
\hline
& $Y=1$ & $Y=2$ & Y=3\\
\hline
$X=1$ & $1 - 3 \epsilon - 4\epsilon^2 - 3 \epsilon^3 - \epsilon^4$ &
$2\epsilon$ & $\epsilon^2$\\
\hline
$X=2$ & $\epsilon$ & $\epsilon^2$ & $2\epsilon^3$\\
\hline
$X=3$ & $2\epsilon^2$ & $\epsilon^3$ & $\epsilon^4$\\
\hline
\end{tabular}
\end{center}
It is easy to check that $X$ and $Y$ are approximately independent.
However, they are not strongly independent.   Suppose, by way of
contradiction, that there exists some probability measure $\nu' \aeq
\nu$ such that $X$ and $Y$ are independent with respect to $\nu'$.  
Note that 
$$\stand{\frac{\nu(X=1 \inter Y=2)}{\nu(X=2 \inter Y=1)}} =
\stand{\frac{\nu(X=3 \inter Y=1)}{\nu(X=1 \inter Y=3)}} =
\stand{\frac{\nu(X=2 \inter Y=3)}{\nu(X=3 \inter Y=2)}} = 2.$$
Since $\nu' \aeq \nu$, it is easy to check that
$$\stand{\frac{\nu'(X=1 \inter Y=2)}{\nu'(X=2 \inter Y=1)}} =
\stand{\frac{\nu'(X=3 \inter Y=1)}{\nu'(X=1 \inter Y=3)}} =
\stand{\frac{\nu'(X=2 \inter Y=3)}{\nu'(X=3 \inter Y=2)}} = 2.$$
Thus, it follows that 
$$\stand{\frac{\nu'(X=1 \inter Y=2) \times \nu'(X=3 \inter Y=1) \times \nu'(X=2
\inter Y=3)}{ \nu'(X=2 \inter Y=1) \times \nu'(X=1 \inter Y=3) \times
\nu'(X=3 \inter Y=2)}} = 8.$$ 
However, since $X$ and $Y$ are independent with respect to $\nu'$, we
must have 
$$\stand{\frac{\nu'(X=1 \inter Y=2) \times \nu'(X=3 \inter Y=1) \times
\nu'(X=2 
\inter Y=3)}{ \nu'(X=2 \inter Y=1) \times \nu'(X=1 \inter Y=3) \times
\nu'(X=3 \inter Y=2) }} = 1.$$
This gives the desired contradiction.
\exam
\commentout{
We can define two events $U$ and $V$ to be strongly independent if the
random variables $\chi_U$ and $\chi_V$ are strongly independent.
However, it is not hard to check (using techniques much like those used
to prove Proposition~\ref{indaeq}) that $\chi_U$ and $\chi_V$ are
strongly independent iff they are weakly (or approximately) independent.
Distinctions that are significant when considering independence of
random variables disappear at the level of independence of independence
of events. 
}
}

The third notion of independence that BBD consider is the strongest.
BBD \citeyear{BBD2} define $X_1, \ldots, X_n$ to be
strongly independent with respect to an LPS 
$\vecmu$ if they are independent (in the usual sense) with respect to an NPS 
$\nu$ such that $\mu \aeq \nu$.%
\footnote{In \cite{BBD2}, BBD say that this definition of strong
independence is given in \cite{BBD1}.  However, the definition appears
to be given only in terms of NPS's in \cite{BBD1}.}
Moreover, they give a characterization
of this notion of strong independence, which I henceforth call \emph{BBD
strong independence}, to distinguish it from the KR notion of strong
independence that I discuss shortly.
Given a tuple $\vec{r} = (r^0, \ldots, r^{k-1})$ of vectors of reals in
$(0,1)^k$ and a finite LPS
$\vecmu = (\mu^0, \ldots, \mu^k)$, let $\vecmu \, \Box \, \vec{r}$
be the (standard) probability measure 
$$(1 - r^0) \mu^0 + r^0[(1-r^1) \mu^1 + r^1[(1-r^2)\mu^2 + r^2[\cdots +
r^{k-2}[(1-r^{k-1})\mu^{k-1} + r^{k-1}\mu^k)]\ldots ]]].$$
Note that $\vecmu \, \Box \, \vec{r}$ is defined only if $\vecmu$ is
finite.  Thus, in discussing BBD strong independence, I restrict  to
finite LPS's.
In addition, for technical reasons that will become clear in the proof
of Theorem~\ref{BBDstrongindependence}, I consider only random variables
with finite range, which is what BBD do as well. 
BBD \citeyear[p.~90]{BBD2} claim without proof that ``it is
straightforward to show'' that $X_1, \ldots, X_n$ are BBD strongly
independent with respect to $\vecmu$ iff there is a
sequence $\vec{r}^j$, $j = 1, 2, \ldots$ of vectors in $(0,1)^k$ 
such that $\vec{r}^j \rightarrow (0,\ldots, 0)$ 
as $j\rightarrow\infty$, 
and $X_1, \ldots, X_n$ are
independent with respect to $\vecmu \, \Box \, \vec{r}^j$ for $j = 1, 2, 3,
\ldots$.  I can prove this result only if the NPS $\nu$ such that
$\vecmu \aeq \nu$ and $X_1, \ldots, X_n$ are independent with respect to
$\nu$ has a range that is an elementary extension of the reals (and thus
has the same first-order properties as the reals).

\thm\label{BBDstrongindependence} 
There exists an NPS $\nu$ whose range  
is an 
elementary extension of the reals such that $\vecmu \aeq \nu$ and $X_1,
\ldots, X_n$ are 
independent with respect to $\nu$ iff there
exists a sequence  
$\vec{r}^j$, $j = 1, 2, \ldots$ of vectors in $(0,1)^k$ 
such that $\vec{r}^j \rightarrow (0,\ldots, 0)$ 
as $j\rightarrow\infty$, 
and $X_1, \ldots, X_n$ are
independent with respect to $\vecmu \, \Box \, \vec{r}^j$ for $j = 1, 2, 3,
\ldots$.   
\ethm
I do not know if this result holds without requiring that $\nu$ be an
elementary extension of the reals.

Kohlberg and Reny \citeyear{KR97} define a notion of strong independence with
respect to what they call {\em relative probability spaces}, which are
closely related to Popper spaces of the form
$(W,2^W,2^W-\{\emptyset\},\mu)$, where all subsets of $W$ are measurable and
it is possible to condition on all nonempty sets.   
Their definition is similar in spirit to the characterization of BBD
strong independence given in Theorem~\ref{BBDstrongindependence}.
For ease of exposition, I recast their definition in terms of Popper spaces.
$X_1, \ldots, X_n$ are {\em KR-strongly independent\/} with respect to the
Popper space $(W,\F,\F', \mu)$, where $\F'$ includes all events of the
form $X_i = x$ for $x \in \V(X_i)$, if there exist a sequence of
standard probability measures $\mu_1, \mu_2, \ldots$ such that $\mu_j
\rightarrow \mu$, and for all $j = 1, 2, 3, \ldots$,
$\mu_j(U) > 0$ for $U \in \F'$ and $X_1,
\ldots, X_n$ are independent with respect to $\mu_j$.
As Kohlberg and Reny show,
KR-strong independence implies approximate independence%
\footnote{They actually show only that it implies weak independence, but
the same argument shows that it implies approximate independence.}
and is, in general, strictly stronger.  

The following theorem characterizes KR strong independence in terms of
NPS's.

\thm\label{KRindependence}  
$X_1, \ldots, X_n$ are KR-strongly independent with respect to the Popper
space $(W,\F,\F',\mu)$ iff there 
exists an NPS $(W,\F,\nu)$ such that 
$\FNP(W,\F,\nu) = (W,\F,\F',\mu)$ and $X_1, \ldots,
X_n$ are independent with respect to $(W,\F,\nu)$.  
\ethm
It follows from the proof that we can require the range of $\nu$ to be a
nonelementary extension of the reals, but this is not necessary.

\commentout{
There is a sense in which KR-strong independence is weaker than
BBD-strong independence.
Define $X_1, \ldots, X_n$ to be KR-strongly independent (resp.,
BBD-strongly independent) with respect to 
NPS $\nu$ if there exists an NPS $\nu'$ such that $\nu \simeq \nu'$
(resp., $\nu \aeq \nu'$) and $X_1, \ldots, X_n$ are independent with
respect to  $\nu'$.  As we have seen, $\simeq$ is a coarser notion of
equivalence than $\sim
}


\commentout{
Now I can compare the definitions given here to those discussed by BBD, 
Hammond, and Kohlberg and Reny.  BBD define a (standard or nonstandard)
probability measure 
$\nu$ on $W= W_1 \times \cdots \times W_n$ to be a product measure if
there exist measures $\nu_i$ on $W_i$ for $i = 1, \ldots, n$, such that
such that $\nu((w_1, \ldots, w_n)) = \nu_1(w_1) \times \cdots \times
\nu_n(w_n)$.  If $X_i$ is the random variable that projects on to the
$i$th component, then it is easy to see that $\nu$ is a product measure
iff $X_1, \ldots, X_n$ are independent.

Hammond mainly focuses on Popper spaces, and follows BBD's lead
in considering when a Popper space can be, in a sense, viewed as a
product measure.   He defines a notion of conditional independence of a 
Popper space defined on $W = W_1 \times \cdots \times W_n$ which is
similar in spirit to the notion of independence of random variables in
Popper spaces as defined here.  In fact, it is straightforward to show
that the Popper space $(W_1 \times \cdots \times W_n, \F,\F',\mu)$ is
conditionally independent in Hammond's sense iff the projections
$X_1, \ldots, X_n$ are independent with respect to the Popper space, in
the sense defined here. }

Kohlberg and Reny show that their notions of weak and strong independence
can be used to characterize
Kreps  and Wilson's
\citeyear{KW82} notion of sequential equilibrium.  
BBD \citeyear{BBD2} use their notion of strong independence in their
characterization of perfect equilibrium and proper equilibrium for games
with more than two players.  
Finally, Battigali \cite{Bat96} uses approximate independence (or,
equivalently, independence in cps's) to characterize sequential
equilibrium. 
\commentout{
Sequential equilibrium uses the notion of an {\em assessment}.
Given a game $\Gamma$, an assessment is 
a pair $(\rho,\pi)$, where $\rho$ is a
function that assigns to each information set $I$ in $\Gamma$ a probability
measure $\rho(I)$ on the set of histories in that information set, and $\pi$
assigns to each a node $x$ a probability $\pi(x)$ on the possible next
moves at that node so that $\pi(x) = \pi(x')$ for two nodes $x$ and $x'$
in the same information set.   Roughly speaking, an assessment is {\em
consistent\/} if,  whenever information set $I$ immediately follows 
information set $I'$, if $I$ can be reached from $I'$ with positive
probability, then  $\rho(I)$ is obtained from
$\rho(I')$ by the obvious computation; if $I$ is not reachable from
$I'$ with positive probability, then $\rho(I)$ must be the
limit of the probabilities on $I$ induced by imposing small trembles on the
moves (so that all of them have some small positive probability, which
goes to 0).  
Kohlberg and Reny \cite{KR97} show that $(\rho,\pi)$ is an
assessment (i.e., $\pi(x) = \pi(x')$ for all $x$ and $x'$ in the same
information set) iff $S_1, \ldots, S_n$ are weakly independent and
$(\rho, \pi)$ is a consistent iff $S_1, \ldots, S_n$ are strongly
independent.  
}

\section{Discussion}\label{discussion}
As the preceding discussion shows, there is a sense in which NPS's
are more general than both Popper spaces and LPS's.  
It would be of interest to get a natural characterization of those NPS's
that are equivalent to Popper spaces and LPS's; this remains an open
problem.  
LPS's are more expressive than Popper measures in finite spaces and in
infinite spaces where we assume countable additivity (in the sense
discussed at the end of Section~\ref{PopperNPS}), but without assuming
countable additivity, they are incomparable, 
as Examples~\ref{counter1} and~\ref{counter2} show.  
Since all of these approaches to representing uncertainty have been
using in characterizing solution concepts in extensive-form games and
notions of admissibility, the results here suggest that it is worth
considering the extent to which these results depend on the particular
representation used.

It is worth stressing here that this notion of equivalence depends on
the fact that I have been viewing cps's, LPS's, and NPS's as
representations of uncertainty.  But, as Asheim \citeyear{Asheim06}
emphasizes, they can also be viewed as representations of conditional
preferences.  Example~\ref{McGee} shows that, even in finite spaces,
NPS's and LPS's can express preferences that cps's cannot.  However, as
Asheim and Perea \citeyear{AP05} point out, in finite spaces, cps's
can also represent conditional  preferences that cannot be represented by
LPS's and NPS's.  See \cite{Asheim06} for a detailed discussion of the
expressive power of these representations with respect to
conditional preferences.

Although NPS's are the most expressive of the three approaches I have
considered,  they have some disadvantages.  In particular,
working with a nonstandard probability measure requires defining and
working with a non-Archimedean field.  
LPS's have the advantage of using just standard probability measures.
Moreover, their lexicographic structure may give useful insights.
It seems to be worth considering the
extent to which LPS's can be generalized so as to increase their
expressive power. 
In particular, it may be of interest to consider LPS's indexed by
partially ordered and not necessarily well-founded sets, rather than
just LPS's indexed by the ordinals.
For example, 
Brandenburger, Friedenberg, and Keisler~\citeyear{BFK04} characterize
$n$ rounds of 
iterated deletion using finite LPS's, for any $n$.
Rather than using a sequence of (finite) LPS's of different lengths to
characterize (unbounded) iterated deletion,  
it seems that a result similar in spirit can be obtained using a single LPS
indexed by the (positive and negative) integers. 

I conclude with a brief discussion of a few other issues raised by this
paper.
\begin{itemize}
\item Belief:
The connections between LPS's, NPS's, and cps's are relevant to the
notion of belief. 
There are two standard notions of belief that can be defined in LPS's.
Say that $U$ is a {\em certain belief\/} in LPS $\vecmu$ of length
$\alpha$ if $\mu_\beta(U) = 1$ for all $\beta < \alpha$; $U$ is {\em
weakly believed\/} if $\mu_0(U) = 1$.  
Brandenburger, Friedenberg, and Keisler \citeyear{BFK04} defined a 
third
notion of belief,
intermediate between weak and strong belief, 
and provided an elegant decision-theoretic justification of it.
According to their definition, an agent {\em assumes $U$ 
in
$\vecmu$\/} if there is some $\beta < \alpha$ such that (a) $\mu_{\beta'}(U) =
1$ for all 
$\beta' \le \beta$, (b) $\mu_{\beta''}(U) = 0$ for all $\beta'' > \beta$, and
(c) $U \subseteq \union_{\beta' \le \beta} \Supp(\mu_{\beta'})$, where
$\Supp(\mu_{\beta'})$ 
denotes the support of 
the probability measure $\mu_{\beta'}$.  (Condition (c) is unnecessary if $W$
is finite, given Brandenburger, Friedenberg, and Keisler's assumption that 
$W = \union_{\beta'} \Supp(\mu_{\beta'})$.)  
There are straightforward analogues of certain belief and weak belief in
Popper spaces.  $U$ is strongly believed in a Popper space
$(W,\F,\F',\mu)$ if $\mu(U \mid V) = 1$ for all $V \in \F'$; $U$ is
weakly believed if $\mu(U \mid V) = 1$ for all $V \in \F'$ such that
$\mu(V) > 0$.  
Analogues of this notion of assumption have been considered elsewhere in
the literature.
Van Fraassen 
\citeyear{vF95} independently defined a 
notion of belief using Popper spaces; in a finite state space, an event
is what van Fraassen calls a \emph{belief core} iff it is assumed in 
the sense of Brandenburger, Friedenberg, 
and Keisler.  Battigalli and Siniscalchi's \citeyear{BS02} notion of
\emph{strong belief} is also essentially equivalent.  
Assumption also corresponds to Stalnaker's \citeyear{Stal98} notion of
\emph{absoutely robust belief} and Asheim and S{\o}vik's \citeyear{AS05}
notion of \emph{robust belief}.
Asheim and S{\o}vik \citeyear{AS05} do a careful comparison of all these
notions (and others).

It is easy to define analogues of certain and weak belief in NPS's:
$U$ is certain belief if $\nu(U) = 1$; $U$ is weakly believed if
$\stand{\nu(U)} = 1$.  
The results of this paper 
suggest that it may also be worth 
investigating an analogue of assumption in NPS's.

\item Nonstandard utility: 
In this paper, while I have allowed probabilities to be
lexicographically ordered or nonstandard, I have implicitly assumed that
utilities are standard real numbers (since I have restricted to
real-valued random variables).
  There is a tradition in decision theory going back to Hausner
\citeyear{Hausner54} and continued recently in a sequence of papers by
Fishburn and Lavalle (see \cite{FL99} and the references therein) 
and Hammond \citeyear{Hammond99} of
considering nonstandard or lexicographically-ordered utilities.  I have
not considered the relationship between these ideas and the ones
considered here, but there may be some fruitful connections.

\item Countable additivity for NPS's:
Countable additivity for standard
probability measures is essentially a continuity condition.  The
fact that $\sum_{i=1}^\infty a_i$ may not be the least upper bound of
the partial sums $\sum_{i=1}^n a_i$ in an NPS leads to a certain lack of
continuity in decision-making.  For example, let $W = \{w_1, w_2, \ldots\}$.
Consider a nonstandard probability measure $\nu$ such that $\nu(w_1) =
1/3 -\epsilon$, $\nu(w_2) = 1/3 + \epsilon$, and $\nu(w_{k+2}) = 1/(3
\times 2^k)$, for $k = 1, 2, \ldots$.  Let $U_n = \{w_3, \ldots, w_n\}$
and let $U_\infty = \{w_3, w_4, \ldots \}$.  Clearly $\nu(U_n) \tendsto
\nu(U_\infty) = 1/3$.  However, $\nu(U_n) < \nu(w_1)$ for all $n$.
Thus, $E_\nu(\chi_{\{w_1\}}) > E_\nu(\chi_{U_n})$ for all $n \ge 3$ although 
$E_\nu(\chi_{\{w_1\}}) < E_\nu(\chi_{U_\infty})$.  

Not surprisingly, the same situations can be modeled with LPS's.
Consider the LPS $(\mu_1, \mu_2)$, where 
$\mu_1 = \stand{\nu}$, $\mu_2(w_1) = 0$, 
$\mu_2(w_2) = 2/3$, and $\mu_2(w_{k+2}) = 1/(3\times 2^k)$ for $k = 1,
2, \ldots$.  It is easy to see
that again $E_{\vecmu}(\chi_{\{w_1\}}) > E_{\vecmu}(\chi_{U_n})$ for all $n
\ge 3$ although  $E_{\vecmu}(\chi_{\{w_1\}}) < E_\nu(\chi_{U_\infty})$.  
(A similar example can be obtained using SLPS's, by replacing each world
$w_i$ by a pair of worlds $w_i', w_i''$, where $w_i'$ is in the support
of $\mu_1$ and $w_i''$ is in the support of $\mu_2$.)

An analogous continuity problem arises even in finite domains.  Let $W = \{w_1, w_2, w_3\}$ and consider a sequence of 
probability measures $\nu_n$ such that $\nu_n(w_1) = 1/3
-1/n$, $\nu_n(w_2) = 1/3 - \epsilon$ and $\nu(w_3) = 1/3 + 1/n +
\epsilon$.  Clearly $\nu_n 
\tendsto \nu$, where $\nu(w_1) = 1/3$, $\nu(w_2) = 1/3 - \epsilon$, and
$\nu(w_3) = 1/3 + \epsilon$.  However, $\nu_n(\chi_{\{w_1\}}) <
\nu_n(\chi_{\{w_2\}})$ for all $n$, while $\nu(\chi_{\{w_1\}}) >
\nu(\chi_{\{w_2\}})$.  Again, the same situation can be modeled using LPS's
(and even SLPS's).

Of course, continuity plays a significant role in standard
axiomatizations of SEU, and is vital in proving the existence of a Nash
equilibrium.  None of the uses of continuity that I am familiar with
have the specific form of this example, but I believe it is worth
considering further the impact of this lack of continuity.
\end{itemize}

\paragraph{Acknowledgments:}  I'd like to thank Adam Brandenburger and
Peter Hammond for a number of very enlightening discussions, Bob
Stalnaker for pointing out Example~\ref{counter1}, Brian Skyrms for
pointing me to Hammond's work, Bas van Fraassen for pointing
me to Spohn's work, Amanda Friedenberg for her careful reading of an
earlier draft, her many useful comments, and for encouraging me to try 
to understand what my results had to say about Battigalli and
Sinischalchi's work, 
and Horacio Arlo-Costa, Geir Asheim, Larry Blume, Adam Brandenburger,
Eddie Dekel, and the anonymous reviewers for a number of
useful comments on earlier drafts of this paper.


\appendix

\section{Appendix: Proofs}
In this section, I prove all the results claimed in the main part of the
paper.  For the convenience of the reader, I repeat the statements of
the results.  

\medskip

\othm{FCPfin} 
If $W$ is finite and $(\F,\F')$, then
$\FCP$ is a bijection from $\SLPS(W,\F,\F')$ to $\Popper(W,\F,\F')$.  
\eothm

\medskip

\prf The first step is to show that $\FCP$ is an injection.  
If $\vecmu, \vecmu' \in \SLPS(W,\F,\F')$ and $\vecmu \ne \vecmu'$, let
$\mu = \FCP(\W,\F,\vecmu)$, and let $\mu' = \FCP(\W,\F,\vecmu')$.    Let
$i$ be the least index such that $\mu_i \ne \mu'_i$.  
There is some set $U$ such that $\mu_i(U) \ne \mu'_i(U)$.  
Let $U_i$ be the set such $\mu_i(U_i) = 1$ and $\mu_j(U_i) = 0$ for $j <
i$; since $\vecmu$ is an SLPS, such a set $U_i$ exists.  Similarly, let
$U_i'$ be such that $\mu_i'(U_i) = 1$ and $\mu_j'(U_i) = 0$ for $j <
i$.  Since $\mu_j = \mu_j'$ for all $j < i$, we must have $\mu_j(U_i \union
U_i') = \mu_j(U_i \union U_i') = 0$ for all $j <  i$.  
Clearly $\vecmu(U_j \union U_j') > 0$, so $U_j \union U_j' \in \F'$.
Moreover,  
$\mu(U \mid U_i \union U_i') = \mu_i(U \mid U_i \union U_i') =
\mu_i(U)$.  Similarly, $\mu'(U \mid U_i \union U_i') = \mu_i'(U)$.
Hence, $\mu \ne \mu'$.

To show that $\FCP$ is a surjection, given a cps $\mu$, let $\vecmu =
(\mu_0, \ldots, \mu_k)$ be the LPS constructed in the main text.  We
must show that  
$\FCP(\vecmu) = (W,\F,\F',\mu)$.  Suppose that  $\FCP(\vecmu) =
(W,\F,\F'',\mu')$. 
I first show that $\F' = \F''$.  Suppose that $V \in \F''$.  Then
$\mu_i(V) > 0$ for some $i$.  Thus, $\mu(V \mid U_i) > 0$.  Since 
$U_i \in \F'$, it follows that $V \in \F'$.  Thus, $\F'' \subseteq \F'$.

To show that $\F' \subseteq \F''$, first note that, by
construction, $\mu(U_j \mid \overline{U_0 \union \ldots \union U_{j-1}}
) = 1$.    
It easily follows that if $V \subseteq \overline{U_0 \union \ldots
\union U_{j-1}}$
then $$\mu(V \mid \overline{U_0 \union \ldots \union U_{j-1}}) = \mu(V
\inter U_j \mid \overline{U_0 \union \ldots \union U_{j-1}}).$$ 
Thus, by CP3,
$$\mu(V \mid \overline{U_0 \union \ldots \union U_{j-1}}) =
\mu(V \inter U_j \mid \overline{U_0 \union \ldots \union U_{j-1}}) = \mu(V \mid
U_j) \times  
\mu(U_j \mid \overline{U_0 \union \ldots \union U_{j-1}}),$$ so 
\begin{equation}\label{eq1}
\mu(V \mid U_j) = \mu(V \mid \overline{U_0 \union \ldots \union U_{j-1}}).
\end{equation}

Now suppose that $V \in \F'$.
Clearly $V \inter (U_0 \union \ldots \union U_k) \ne
\emptyset$, for otherwise $V \subseteq \overline{U_0 \union \ldots
\union U_k}$, contradicting the fact that $\overline{U_0 \union \ldots
\union U_k} \notin \F'$.  Let $j_V$ be the smallest index $j$ such that $V
\inter U_j \ne \emptyset$. 
I claim that $\mu(V \mid \overline{U_0 \union \ldots \union U_{j_V - 1}}) \ne
0$.  For if $\mu(V \mid \overline{U_0 \union \ldots \union U_{j_V - 1}}) = 
0$, then $\mu(U_{j_V} - V \mid \overline{U_0 \union \ldots \union U_{j_V -
1}}) = 1$, contradicting the definition of $U_{j_V}$ 
as the smallest set $U'$ such that $\mu(U' \mid \overline{U_0 \union
\ldots \union U_{j_V - 1}}) = 1$.    Moreover, since 
$V \subseteq \overline{U_0 \union \ldots U_{j_V-1}}$, it follows
from (\ref{eq1}) that 
$\mu(V \mid U_{j_V}) = \mu(V \mid \overline{U_0 \union
\ldots \union U_{j_V - 1}}) > 0$.  Thus, $\mu_{j_V}(V) > 0$, so $V \in
\F''$.  

This argument can be extended to show that $\mu(V' \mid V) = \mu'(V' \mid
V)$ for all $V' \in \F$.
Since $V \inter U_i = \emptyset$ for $i < j_V$, it follows that
$\mu'(V' \mid V) = \mu_{j_V}(V' \mid V)$.  
By CP3, $\mu(V' \mid V) \times \mu(V \mid \overline{U_0 \union \ldots
\union U_{j_V 
- 1}}) = \mu(V'\inter V \mid \overline{U_0 \union \ldots \union U_{j_V - 1}})$.
By (\ref{eq1}) and the fact that $\mu(V \mid U_{j_V}) > 0$, it follows
that $\mu(V' \mid V) = \mu(V'\inter V \mid U_{j_V})/\mu(V \mid
U_{j_V})$, 
that is,  that $\mu(V' \mid V) = \mu_{j_V}(V' \mid V)$.
\eprf

\bigskip

Although Theorem~\ref{infiso} was proved by Spohn \citeyear{Spohn86}, I
include a proof here as well, to make the paper self-contained.

\othm{infiso} For all $W$, the map $\FCP$ is a bijection from 
$\SLPS^c(W,\F,\F')$
to $\Popper^c(W,\F,\F')$.  \eothm

\medskip

\prf Again, the difficulty comes in showing that $\FCP$ is onto.  
As it says in the main text, given a Popper space $(W,\F,\F',\mu)$, the
idea is to
construct sets $U_0, U_1, \ldots$ and an LPS $\vecmu$ such that
$\mu_\beta(V)=\mu(V \mid U_\beta)$, and show that $\FCP(W,\F,\vecmu) =
(W,\F,\F',\mu)$. The construction is somewhat involved.

As a first step, put an order $\le$ on sets in
$\F'$ by defining $U \le V$ if 
$\mu(U \mid U \union V) > 0$.  
(Essentially, the same order is considered by van Fraassen \citeyear{vF76}.)

\lem\label{lem0}  $\le$ is transitive.  \elem

\prf 
By definition, if $U \le V$ and $V \le V'$, then $\mu(U \mid U \union V) >0$ 
and $\mu(V \mid V \union V') > 0$.  To see that $\mu(U \mid U \union
V') > 0$, note that
$\mu(U \mid U \union V \union V') + \mu(V \mid U \union V \union V') + \mu(V' \mid U
\union V \union V') = 1$, so at least one of $\mu(U \mid U \union V \union
V')$, $\mu(V \mid U \union V \union V')$, or $\mu(V' \mid U \union V \union V')$
is positive.  I consider each of the cases separately.

\paragraph{Case 1:} Suppose that $\mu(U \mid U \union V \union V') > 0$.  By CP3, 
$$\mu(U \mid U \union V \union V') = \mu(U \mid  U \union V') \times \mu(U \union
V' \mid U \union V \union V').$$
Thus, $\mu(U \mid U \union V') > 0$, as desired.

\paragraph{Case 2:} Suppose that $\mu(V \mid U \union V \union V') > 0$.  
By assumption, $\mu(U \mid U \union V) > 0$; since $\mu(V \mid U \union V \union
V') > 0$, it follows that $\mu(U \union V \mid U \union V \union V') > 0$.
Thus, by CP3, 
$$\mu(U \mid U \union V \union V') = \mu(U \mid  U \union V) \times \mu(U \union
V \mid U \union V \union V') > 0.$$
Thus,  case 2 can be reduced to case 1.

\paragraph{Case 3:} Suppose that $\mu(V' \mid U \union V \union V') > 0$.  
By assumption, $\mu(V \mid V \union V') > 0$; since $\mu(V' \mid U \union V \union
V') > 0$, it follows that $\mu(V \union V' \mid U \union V \union V') > 0$.
Thus, by CP3, 
$$\mu(V \mid U \union V \union V') = \mu(V \mid  V \union V') \times \mu(V \union
V' \mid U \union V \union V') > 0.$$
Thus, case 3 can be reduced to case 2.

This completes the proof, showing that $\le$ is transitive.
\eprf

Define  $U \sim V$ if $U \le V$ and $V \le U$.   

\lem\label{lem1} $\sim$ is an equivalence relation on $\F'$. \elem

\prf It is immediate from the
definition that $\sim$ is reflexive and symmetric; transitivity follows
from the transitivity of $\le$.   \eprf

R\'{e}nyi \citeyear{Renyi56}
and van Fraassen \citeyear{vF76} also considered the $\sim$ relation in
their papers, and the argument that $\le$ is transitive is similar in
spirit to R\'{e}nyi's argument that $\sim$ is transitive.
However, the rest of this proof diverges from those of R\'{e}nyi and van
Fraassen. 

Let $[U]$ denote the $\sim$-equivalence class of $U$, and 
let $\F'/\nsim = \{[U]: U \in \F'\}$.

\lem\label{lem2} Each equivalence class $[V] \in \F'/\nsim$ is closed under
countable unions.  \elem

\prf Suppose that $V_1, V_2, \ldots \in [V]$.  I must show that 
$\union_{i=1}^\infty V_i \in [V]$.  Clearly $V_j \le \union_{i=1}^\infty
V_i$ for all $j$.  Suppose, by way of contradiction, that 
$\union_{i=1}^\infty V_i \not\le V_j$ for some $j$.  Since $\le$ is
transitive, it follows that $V_j < \union_{i=1}^\infty V_i$ for all $j$.
Thus, $\mu(V_j \mid \union_{i=1}^\infty V_i) = 0$ for all $j$.
But then, by countable additivity,
 $$1 = \mu(\union_{i=1}^\infty V_i \mid \union_{i=1}^\infty V_i) \le
\sum_{j=1}^\infty \mu(V_j \mid \union_{i=1}^\infty V_i) = 0,$$
a contradiction.  Thus, $[V]$ is closed under countable unions.
\eprf

\commentout{
Next, observe that we can define a total preorder $\preceq$ (i.e., a
reflexive and transitive relation) on $[V]$ using
the same techniques as used to define $\le$.  Namely, 
$V_1 \preceq V_2$ if $\mu(V_1 \mid V_1 \union V_2) \le \mu(V_2 \mid V_1 \union
V_2)$.  To see that $\preceq$ is transitive, suppose that $V_1, V_2, V_3
\in [V]$ and $V_1 \preceq V_2$ and $V_2 \preceq V_3$.  
By CP3, 
\begin{equation}\label{eq2}
\mu(V_i \mid V_1 \union V_2 \union V_3) = \mu(V_i \mid V_i \union V_j)
\times \mu(V_i \union V_j \mid V_1 \union V_2 \union V_3),
\end{equation}
for all $i, j$. 
Since $V_1 \le V_2, applying (\ref{eq2}) first with $i = 1$ and $j=2$
and then with $i=2$ and $j=1$, 
it follows that $\mu(V_1 \mid V_1 \union V_2 \union V_3)
\le \mu(V_2 \mid V_1 \union V_2 \union V_3)$.  Similarly, since $V_2 \le
V_3$, it follows that $\mu(V_2 \mid V_1 \union V_2 \union V_3)
\le \mu(V_3 \mid V_1 \union V_2 \union V_3)$.  
Thus, 
\begin{equation}\label{eq3}
\mu(V_1 \mid V_1 \union V_2 \union V_3) \le \mu(V_3 \mid V_1 \union V_2 \union
V_3).
\end{equation}
Since $[V]$ is
closed under unions, it follows that $V_1 \union V_3 \in [V]$ and
$V_1 \union V_2 \union V_3 \in [V]$.  Thus, $\mu(V_1 \union V_3 \mid V_1
\union V_2 \union V_3) > 0$.  Now it immediately follows from
(\ref{eq2}) and (\ref{eq3}) that $\preceq$ is transitive.}

Fix an element $V_0 \in [V]$.  
\lem\label{lem3} $\inf \{\mu(V_0 \mid V_0 \union V'): V' \in [V]\} > 0$. \elem

\prf Suppose that $\inf \{\mu(V_0 \mid V_0 \union V'): V' \in [V]\} = 0$.
Then there exist sets $V_1, V_2, \ldots$ such that $\mu(V_0 \mid V_0 \union
V_n) < 1/n$.  Since $[V]$ is closed under countable unions,
$\union_{i=1}^n V_i \in [V]$.  Since $V_0 \sim \union_{i=1}^n V_i$, it
follows that $\mu(V_0 \mid \union_{i=0}^\infty V_i) > 0$.
But, by CP3, $$\mu(V_0 \mid \union_{i=0}^\infty V_i) = \mu(V_0 \mid V_0 \union V_n)
\times \mu(V_0 \union V_n \mid \union_{i = 0}^\infty V_i) \le \mu(V_0 \mid V_0
\union V_n) \le 1/n.$$
Since this is true for all $n > 0$, it follows that 
$\mu(V_0 \mid \union_{i=0}^\infty V_i) = 0$, a contradiction.
\eprf

The next lemma shows that each equivalence class in $\F'/\nsim$ has a
``maximal element''. 

\lem\label{lem4} In each equivalence class $[V]$, there is an element
$V^*\in [V]$ such that $\mu(V^* \mid V' \union V^*) = 1$ for all $V' \in
[V]$. \elem 

\prf Again, fix an element $V_0 \in [V]$.  By Lemma~\ref{lem3}, there
exists some $\alpha_V > 0$ such that $\inf \{\mu(V_0 \mid V_0 \union V'): V'
\in [V]\} = \alpha_V$. Thus, there exist sets $V_1, V_2, V_3, \ldots \in
[V]$ such that $\mu(V_0 \mid V_0 \union V_n) < \alpha + 1/n$.  By
Lemma~\ref{lem2}, $V^* = \union_{i=0}^\infty V_i \in [V]$.   By CP3,
$$\mu(V_0 \mid V^*) = \mu(V_0 \mid V_0 \union V_n) \times \mu(V_0 \union V_n \mid V^*)
\le \mu(V_0 \mid V_0 \union V_n) < \alpha_V + 1/n.$$
Thus, $\mu(V_0 \mid V^*) \le \alpha_V$.  By choice of $\alpha_V$, it follows
that $\mu(V_0 \mid V^*) = \alpha_V$.

Suppose that
$\mu(V^* \mid V' \union V^*) < 1$ for some $V' \in [V]$.  But then, by CP3,
$$\mu(V_0 \mid V' \union V^*) = \mu(V_0 \mid V^*) \times \mu(V^* \mid V' \union V^*) <
\alpha_V,$$ contradicting the choice of $\alpha_V$.  Thus,  
$\mu(V^* \mid V' \union V^*) = 1$ for all $V' \in [V]$.  \eprf

Define a
total order on these equivalence relations by taking $[U] \le [V]$ if
$U' \le V'$ for some $U' \in [U]$ and $V' \in [V]$.  It is easy to check
(using the transitivity of $\le$) that if $U' \le V'$ for some $U' \in
[U]$ and some $V' \in [V]$, then $U'' \le V''$ for all $U'' \in [U]$ and
all $V'' \in [V]$.   

\lem $\le$ is a well-founded relation on $\F'/\nsim$. \elem

\prf
Note that if $[U] < [V]$, then $\mu(V \mid U \union V) = 0$.  It now
follows from countable additivity that $<$ is a well-founded order on
these equivalence classes.  For suppose that there exists an infinite 
decreasing sequence $[U_0] > [U_1] > [U_2] > \ldots$.  
Since $\F$ is a $\sigma$-algebra, $\union_{i=0}^\infty U_i \in \F$; since
$\F'$ is closed under supersets, $\union_{i=0}^\infty U_i \in \F'$.  
By CP3,
$$\mu(U_j \mid \union_{i=0}^\infty U_i) = \mu(U_j \mid U_{j} \union U_{j+1}) \times
\mu(U_j \union U_{j+1} \mid \union_{i=0}^\infty U_i) = 0.$$  Let $V_0 = U_0$
and, for $j > 0$, let
$V_j = U_j - (\union_{i =0}^{j-1} U_j)$.  Clearly the $V_j$'s are
pairwise disjoint, $\union_i U_i = \union_i V_i$, and
$\mu(V_j \mid \union_{i=0}^\infty U_i) \le \mu(U_j \mid \union_{i=0}^\infty U_i) = 0$.  
It now follows that using countable additivity that 
$$1 = \mu(\union_{i=0}^\infty U_i \mid \union_{i=0}^\infty U_i) =
\sum_{i=0}^\infty \mu(V_i \mid  \union_{i=0}^\infty U_i) = 0.$$
This is as contradiction, so the equivalence classes are well-founded.
\eprf

Because $\le$ is well-founded, there is an order-preserving bijection
$O$ from $\F'/\nsim$ to an initial segment of the ordinals (i.e., $[U]
\le [V]$ iff $O([U]) \le O([V])$.  
Thus, the equivalence classes can be enumerated using all the ordinals
less than some ordinal $\alpha$.  By Lemma~\ref{lem4}, there are
sets $U_\beta$, $\beta < \alpha$, in $\F'$ such that if $O([U]) =
\beta$, then $U_\beta \in [U]$ and $\mu(U_\beta \mid U \union U_\beta) = 1$
for all $U' \in [U]$.  Define an LPS $\vecmu = (\mu_0, \mu_1, \ldots )$ of
length $\alpha$ by taking $\mu_\beta(V) = \mu(V \mid U_\beta)$.  The choice
of the $U_\beta$'s guarantees that this is actually an SLPS.

It remains to show that $(W,\F,\F',\mu)$ is the result of applying 
$\FCP$  
to $(W,\F,\vecmu)$.  Suppose that instead $(W,\F,\F'',\mu')$ is the
result.  The argument that $\F'' \subseteq \F'$ is identical to that in
the finite case: If $V \in \F''$, then
$\mu_\beta(V) > 0$ for some $\beta$.  Thus, $\mu(V \mid U_\beta) > 0$.  Since 
$U_\beta \in \F'$, it follows that $V \in \F'$.  Thus, $\F'' \subseteq \F'$.

Now suppose that $V \in \F'$.  Thus, $V \sim V_\beta$ for some $\beta <
\alpha$.  It follows that $\mu(V \mid V_\beta) > 0$, so $V \in \F''$.

Finally, to show that $\mu(U \mid V) = \mu'(U \mid V)$, suppose that $\beta$ is
such that $V \sim V_\beta$.  It follows that $\mu(V \mid V_{\beta'}) = 0$ for
$\beta' < \beta$ and $\mu(V \mid V_{\beta}) > 0$.  Thus, by definition,
$\mu'(U \mid V) = \mu_\beta(U \mid V)$.  Without loss of generality, assume
that $U \subseteq V$ (otherwise replace $U$ by $U \inter V$).  Thus, by
CP3, 
\begin{equation}\label{eq4}
\mu(U \mid V) \times \mu(V \mid V \union V_\beta) = \mu(U \mid V \union V_\beta).
\end{equation}
Suppose $V' \subseteq V$.
Clearly $$\mu(V' \mid V \union V_\beta) = \mu(V' \inter V_\beta \mid V \union
V_\beta) + \mu(V' \inter \overline{V_\beta} \mid V \union V_\beta).$$
Now by CP3 and the fact that  $\mu(V_\beta \mid V \union V_\beta) = 1$,
$$\mu(V' \inter V_\beta \mid V \union V_\beta) = \mu(V' \mid V_\beta) \times
\mu(V_\beta \mid V \union V_\beta) = \mu(V' \mid V_\beta)$$
and 
$$\mu(V' \inter \overline{V_\beta} \mid V \union V_\beta) \le
\mu(\overline{V_\beta} \mid V \union V_\beta) = 0.$$
Thus, $\mu(V' \mid V \union V_\beta) = \mu(V' \mid V_\beta)$.
Applying this observation to both $U$ and $V$ shows that
$\mu(V \mid V \union V_\beta) = \mu(V \mid V_\beta)$ and $\mu(U \mid V
\union V_\beta) 
=\mu(U \mid V_\beta)$.  Plugging this into (\ref{eq4}), it follows that
$$\mu(U \mid V) = \mu(U \mid V_\beta)/\mu(V \mid V_\beta) = \mu_\beta(U)/\mu_\beta(V) =
\mu_\beta(U \mid V) = \mu'(U \mid V).$$
This completes the proof of the theorem.
\eprf

\bigskip
\opro{prop:BS} The map $\FCP$ is a surjection from 
$\SLPS^c(W,\F,\F')$ onto $\T^c(W,\F,\F')$.  \eopro

\medskip

\prf Suppose that $\mu \in \T^c(W,\F,\F')$.  I want to construct an
SLPS $\vecmu \in \SLPS^c(W,\F,\F')$ such that $\FCP(\vecmu) = \mu$. 
I first label each element of $\F'$ with a natural
number.  Intuitively, if $U \in \F'$ is labeled $k$, then $k$ will be
the least index such that $\mu_k(U) > 0$.  The labeling is done by
induction on $k$.  Each topmost set in the forest
(i.e., the root of some tree in the forest) is labeled 0, as are all
sets $U'$ such that $\mu(U' \mid U) > 0$, where $U$ is a topmost node.
These are all the nodes labeled by 0.  Label all the maximal unlabeled
sets by 1 (that is, label $U \in \F'$ by 1 if it is not labeled 0, and
is not a subset of another unlabeled set); in addition, label a set $U'$
by 1 if $\mu(U' \mid U) > 0$ and $U$ is labeled by 1.  Note that every
set at depth 0 or 1 in the forest is labeled by either 0 or 1.

Suppose that the labeling process has been completed for labels $0,
\ldots, k$ such that the following properties hold, where $\lab(U)$
denotes the label of the event $U$:  
\begin{itemize}
\item all sets up to depth $k$ in the forest have been labeled;
\item if $\lab(U) = k'$, $U' \in \F'$, and $\mu(U' \mid U) > 0$, then
$\lab(U') \le \lab(U)$.  
\end{itemize}
Label all the maximal unlabeled sets with $k+1$; in addition, if $U'$
is unlabeled and $\mu(U' \mid U) > 0$ for some $U$ such that $\lab(U)
= k+1$, then assign label $k+1$ to $U'$.  Clearly the two properties
above continue to hold.  This completes the labeling process.

Let $\C_k$ be the set of maximal sets in $\F'$ labeled $k$.  
T2 and T3 guarantee that, for all $k$, the sets in $\C_k$ are
disjoint.  Let $\mu_k'$ be 
an arbitrary probability on $W$ such that $\mu_k'(U) > 0$ for all $U \in
\C_k$ and $\sum_{U \in C_k} \mu_k'(U) = 1$.  Define an LPS 
$\vecmu = (\mu_0, \mu_1, \ldots)$ as follows (where the length of
$\vecmu$ is $\omega$ if $\C_k \ne \emptyset$ for all $k$, and 
is $k+1$ if $k$ is the largest integer such that $\C_k \ne \emptyset$).
For $V \in \F$, let $\mu_j(V) = \sum_{U \in \C_j} \mu(V \mid U)
\mu_j'(U)$.  I now show that $\vecmu(V \mid U) = \mu(V \mid U)$ for all $V \in
\F$ and $U \in \F'$.  Suppose that $U \in \C_k$.  Then $\mu_j(U) = 0$ for
all $j < k$, and $\mu_k(U) > 0$.  Thus, $\vecmu(V \mid U) = \mu_k(V \mid
U)$.  But it is immediate from the definition that $\mu_k(V \mid U) =
\mu(V \mid U)$.  Thus, $\FCP(\vecmu) = \mu$.  Moreover, if $U \in \F'$
and $\lab(U) = k$, let $U'$ be the maximal set containing $U$ such
that $\lab(U') = k$.  (The labeling guarantees that such a set
exists.)  Then $\mu_k(U') = \mu(U' \mid U) > 0$.  It follows that
$\vecmu(U) > 0$ for all $u \in \F'$.  Finally, note that 
$\vecmu$ is an SLPS (in fact, an LCPS). If $U_k = \union \C_k -
\union_{k' > k} (\union \C_{k'})$, then the sets $U_k$ are disjoint, and
$\mu_k(U_k) = 1$.  \eprf

\bigskip

\opro{FCPaeq} 
If $\nu \aeq \vecmu$, then $\nu(U) > 0$ iff $\vecmu(U) > \vec{0}$.
Moreover, if $\nu(U) > 0$, then $\stand{\nu(V \mid U)} = \mu_j(V \mid U)$, where
$\mu_j$ is the first probability measure in $\vecmu$ such that $\mu_j(U)
> 0$. \eopro

\medskip

\prf  Recall that for  $U \subseteq W$, $\chi_U$ is the indicator
function for $U$; 
that is, $\chi_U(w) = 1$ if $w \in U$ and $\chi_U(w) = 0$ otherwise.
Notice that $E_\nu(\chi_U) > E_\nu(\chi_{\emptyset})$ iff $\nu(U) > 0$
and $E_{\vecmu}(\chi_U) > E_{\vecmu}(\chi_{\emptyset})$ iff $\vecmu(U) >
\vec{0}$.  Since $\nu \aeq \vecmu$, it follows that 
$\nu(U) > 0$ iff $\vecmu(U) > \vec{0}$.  If $\nu(U) > 0$, 
note that $E_\nu(\chi_{U\inter V} - r \chi_U) >
E_\nu(\chi_{\emptyset})$ iff $r < \stand{\nu(V \mid U)}$.  Similarly,
$E_{\vecmu}(\chi_{U\inter V} - r \chi_U) >
E_{\vecmu}(\chi_{\emptyset})$ iff $r < \mu_j(U)$, where $j$ is the
least index such that $\mu_j(U) > 0$.  It follows that $\stand{\nu(V \mid U)}
= \mu_j(V \mid U)$.  \eprf

\bigskip

\opro{motivation} If $\vecmu, \vecmu' \in \SLPS(W,\F)$, then
$\vecmu \aeq \vecmu'$ 
iff $\vecmu = \vecmu'$.  
\eopro

\medskip

\prf Clearly $\vecmu = \vecmu'$ implies that $\vecmu \aeq \vecmu'$.
For the converse, suppose that $\vecmu \aeq \vecmu'$ for $\vecmu,
\vecmu' \in \SLPS(W,\F)$.  If $\vecmu \ne \vecmu'$, let $\alpha$ be the
least ordinal such that $\mu_\alpha \ne \mu'_\alpha$, and let $U$ be
such that $\mu_\alpha(U) \ne \mu'_\alpha(U)$.  Without loss of
generality, suppose that $\mu_\alpha(U) >  \mu'_\alpha(U)$.  
Let the sets $U_\beta$
be such that $\mu_\beta(U_\beta) = 1$ and $\mu_\beta(U_\gamma) = 0$ if
$\gamma > \beta$; similarly choose the sets $U_\beta'$.  Since
$\mu_\beta = \mu'_\beta$ for $\beta < \alpha$, it follows that
$\mu_\beta(U_\alpha \union U'_\alpha) = \mu'_\beta(U_\alpha \union
U'_\alpha) = 0$ for $\beta < \alpha$; moreover,
$\mu_\alpha(U_\alpha \union U'_\alpha) = \mu_\alpha'(U_\alpha \union
U'_\alpha) = 1$.  Choose $r$ such that $\mu_\alpha(U) > r >
\mu'_\alpha(U)$.  Let $X$ be the random variable $\chi_U -
r\chi_{U_\alpha \union U'_\alpha}$ and let $Y = \chi_\emptyset$.
Then $E_{\vecmu}(X) > E_{\vecmu}(Y)$, while 
$E_{\vecmu'}(X) < E_{\vecmu'}(Y)$, so $\vecmu \not\aeq \vecmu'$.
\eprf

\bigskip

\opro{finiteeq} If $W$ is finite, then every LPS over $(W,\F)$ is
equivalent to an LPS of length at most $|\Bas(\F)|$. \eopro

\medskip

\prf Suppose that $W$ is finite and $\Bas(\F) = \{U_1, \ldots, U_k\}$.
Given an LPS $\vecmu$, define a finite subsequence $\vecmu' =
(\mu_{k_0}, \ldots, \mu_{k_h})$ of
$\vecmu$  as follows.  Let $\mu_{k_0} = \mu_0$.  Suppose that 
$\mu_{k_0}, \ldots, \mu_{k_j}$ have been defined.  If all probability
measures in $\vecmu$ 
with index greater that $k_j$ are linear combinations of the probability
measures with index $\mu_{k_0}, \ldots, \mu_{k_j}$, then take $\vecmu'
= (\mu_{k_0}, \ldots, \mu_{k_j})$.  Otherwise, let $\mu_{k_{j+1}}$ be
the probability measure in $\vecmu$ with least index that is not a
linear combination of $\mu_{k_0}, \ldots, \mu_{k_j}$.  
Since a probability measure over $(W,\F)$ is determined by its value on
the sets in $\Bas(\F)$,  
a probability measure over $(W,\F)$ can be identified with a vector in
$\IR^{|\Bas(\F)|}$: the vector defining the probabilities of the
elements in $\Bas(\F)$.  There can be at most $|\Bas(\F)|$ linearly
independent such vectors, thus $\vecmu'$ has length at most
$|\Bas(\F)|$.

It remains to show that $\vecmu'$ is equivalent to $\vecmu$.  Given
random variables $X$ and $Y$, suppose that $E_{\vecmu}(X) <
E_{\vecmu}(Y)$.  Then there is some minimal index $\beta$ such that
$E_{\mu_\gamma}(X) = E_{\mu_\gamma}(Y)$ for all $\gamma < \beta$ and
$E_{\mu_\beta}(X) < E_{\mu_\beta}(Y)$.  It follows that 
$\mu_\beta$ cannot be a linear combination of $\mu_\gamma$ for $\gamma <
\beta$.  Thus, $\mu_\beta$ is one of the  probability measures in
$\vecmu'$.  Moreover, the expected value of $X$ and $Y$ agree for all
probability measures in $\vecmu'$ with lower index (since they do in
$\vecmu$).  Thus, $E_{\vecmu'}(X) < E_{\vecmu'}(X)$.  

The argument in the other direction is similar in spirit and left to the
reader.  \eprf

\othm{lpsnps} If $W$ is finite, then 
$\FLN$ is a bijection from $\LPS(W,\F)/\naeq$ to $\NPS(W,\F)/\naeq$
that preserves equivalence (that is, each NPS in $\FLN([\vecmu])$ is
equivalent to $\vecmu$).
\eothm

\prf   I first provide a sufficient condition for an NPS to be equivalent
an LPS in a finite space.

\lem\label{aeqchar} Suppose that $\vecmu = (\mu_0,\ldots, \mu_k)$, and
$\epsilon_0, \ldots, \epsilon_k$ are
such that $\stand{\epsilon_{i+1}/\epsilon_{i}} = 0$ for $i = 1, \ldots,
k-1$ and $\sum_{i=0}^k \epsilon_i = 1$.  Then $\vecmu \aeq \epsilon_0
\mu_0 + \cdots + \epsilon_k       
\mu_k$.%
\footnote{Although I do not need this fact here, it is easy to see that
if $W$ is finite and $\vecmu = (\mu_0, \ldots, \mu_k)$ is 
an SLPS in $\LPS(W,\F)$, then the converse of Lemma~\ref{aeqchar} holds
as well: if $\nu \aeq \vecmu$, then  $\nu = \epsilon_0 \mu_0 + \cdots
\epsilon_k \mu_k$ for some $\epsilon_0, \ldots, \mu_k$ are such that
$\stand{\epsilon_{i+1}/\epsilon_{i}} = 0$ for $i = 1, \ldots,
k-1$ and $\sum_{i=0}^k \epsilon_i = 1$.  (I conjecture this fact is true
in general, not just if $\vecmu$ is an SLPS, but I have not checked this.}
\elem 

\prf Suppose that there exist $\epsilon, \ldots, \epsilon_k$ as in the
statement of the lemma and
$\nu = \epsilon_0 \mu_0 + \cdots + \epsilon_k \mu_k$.  I want to show
that $\vecmu \aeq \nu$.

If $E_{\vecmu}(X) < E_{\vecmu}(Y)$,
then there exists some $j \le k$ such 
that $E_{\mu_j}(X) < E_{\mu_j}(Y)$ and $E_{\mu_{j'}}(X) =
E_{\mu_{j'}}(Y)$ for all $j' < j$.  
Since $E_\nu(X) = \sum_{i=0}^k \epsilon_i E_{\mu_i}(X)$ and 
$E_\nu(Y) = \sum_{i=0}^k \epsilon_i E_{\mu_i}(Y)$, 
to show that $E_\nu(X) < E_\nu(Y)$, it suffices to show
that $\epsilon_j(E_{\mu_j}(Y) - E_{\mu_j}(X)) > 
\sum_{i=j+1}^k \epsilon_i (E_{\mu_i}(X) - E_{\mu_i}(Y))$.  Since
$\epsilon_{j'+1} \le \epsilon_{j'}$ for $j' \ge j$
(this follows from the fact that $\stand{\epsilon_{j'+1}/\epsilon{j'}} =
0$), it follows that 
$\sum_{i=j+1}^k \epsilon_i (E_{\mu_i}(X) - E_{\mu_i}(Y)) \le
\epsilon_{j+1} \sum_{i=j+1}^k |E_{\mu_i}(X) - E_{\mu_i}(Y)|$.
Thus, it suffices to show that $\epsilon_{j+1} \sum_{i=j+1}^k
|E_{\mu_i}(X) - E_{\mu_i}(Y)| <  \epsilon_j(E_{\mu_j}(Y) -
E_{\mu_j}(X))$.  
This is trivially the case if $E_{\mu_i}(X) = E_{\mu_i}(Y)$ for all 
$i$ such that $j+1 \le i \le k$.  Thus, assume without loss of
generality that $\sum_{i=j+1}^k |E_{\mu_i}(X) - E_{\mu_i}(Y)| > 0$.
In this case, it suffices to show that $\epsilon_{j+1}/\epsilon_{j} <
(E_{\mu_j}(Y) - E_{\mu_j}(X))/\sum_{i=j+1}^k |E_{\mu_i}(X) -
E_{\mu_i}(Y)|$. Since the right-hand side of the inequality is  a
positive real and $\stand{\epsilon_{j+1}/\epsilon_{j}} = 0$, the result
follows.

The argument in the opposite direction is similar.  Suppose that
$E_\nu(X) < E_\nu(Y)$.  
Again, since $E_\nu(X) = \sum_{i=0}^k \epsilon_i
E_{\mu_i}(X)$ and  $E_\nu(Y) = \sum_{i=0}^k \epsilon_i E_{\mu_i}(Y)$, 
it must be the case that if  $j$ is the least index such that
$E_{\mu_j}(X) \ne E_{\mu_j}(Y)$, then  $E_{\mu_j}(X) < E_{\mu_j}(Y)$.
Thus, $E_{\vecmu}(X) < E_{\vecmu}(Y)$.  It follows that $\vecmu \aeq \nu$.
\eprf

It remains to show that, given an NPS $(W,\F,\nu)$, there is an equivalence
class $[\vecmu]$ such that $\FLN([\vecmu]) = [\nu]$.  
As I said in the main text, the goal now is to find (standard) probability
measures  
$\mu_0, \ldots, \mu_{k}$ and $\epsilon_0, \ldots, \epsilon_k$ such that
$\stand{\epsilon_{i+1}/\epsilon_i} = 0$ and $\nu = \epsilon_0 \mu_0 +
\cdots + \epsilon_k\mu_k$.  If this can be done then, by
Lemma~\ref{aeqchar}, $\nu \aeq (\mu_0, \ldots, \mu_k)$, and we are done.

Suppose that $\Bas(\F) = \{U_1, \ldots, U_k\}$
and that $\nu$ has range $\IR^*$.  Note that 
a probability measure $\nu'$ on $\F$ can be identified with a
vector $(a_1, \ldots, a_k)$ over $\IR^*$, where $\nu'(U_i) = a_i$, so
that $a_1 + \cdots + a_k = 1$.  In the rest of this proof, I frequently
identify $\nu$ with such a vector.

\lem\label{newlem1} There exist $k' \le k$, $\epsilon_0, \ldots,
\epsilon_{k'}$ where $\epsilon_0 = 1$, 
$\stand{\epsilon_{i+1}/\epsilon_{i}} = 0$ for $i =
1, \ldots, k'-1$, and standard real-valued vectors 
$\vec{b}_j$, $j = 0, \ldots, k'$, in $\IR^k$ such that
$$\nu = \sum_{j=0}^{k'} \epsilon_j \vec{b}_j.$$
\elem

\prf I show by induction on $m \le k$ that there exist $\epsilon_0,\ldots,
\epsilon_m$ and $m' \le m$ such that $\epsilon_j = 0$ for $j' > m'$,
$\stand{\epsilon_{i+1}/\epsilon_{i}} = 0$ for $i =
1, \ldots, m'-1$, and standard vectors 
$\vec{b}_j$ $j = 0, \ldots, m-1$ 
and a possibly nonstandard vector $\vec{b}'_m = 
(b'_{m1}, \ldots, b'_{mk})$ such that
(a) $\nu = \sum_{j=0}^{m-1} \epsilon_j \vec{b}_j + \epsilon_m \vec{b}'_m$,
(b) $|b'_{mi}| \le 1$, and (c) at least $m$ of $b'_{m1}, \ldots, b'_{mk}$
are standard.

For the base case (where $m=0$), just take $\vec{b}'_0 = 
\nu$ and $\epsilon_0 = 1$.  For the inductive step, suppose that $0 \le m
< k$. If  $\vec{b}'_m$ is standard, then take $\vec{b}_m = \vec{b}'_m$,
$\vec{b}_{m+1} = \vec{0}$, and $\epsilon_{m+1}
= 0$.  Otherwise, let $\vec{b}_m = \stand{\vec{b}'_m}$ and let
$\vec{b}''_{m+1} = 
\vec{b}'_m - \vec{b}_m$.  Let $\epsilon' = \max\{|b''_{(m+1)i}|: i = 1,
\ldots, k\}$.    Since not all components of $\vec{b}'_m$ are standard,
$\epsilon' > 0$.  Note that, by construction, $\stand{\epsilon'/ b_{mi}} =
0$ if $b_{mi} \ne 0$, for $i = 1, \ldots, k$.  Let $\vec{b}'_{m+1} =
\vec{b}''_{m+1}/\epsilon'$ and let $\epsilon_{m+1} = \epsilon'
\epsilon_m$.
By construction, $|b'_{(m+1)i}| \le 1$ and at least one
component of $\vec{b}'_{m+1}$ is either 1 or $-1$.  Moreover, if
$b_{mi}'$ is standard, then $b''_{(m+1)i} = b'_{(m+1)i} = 0$.  Thus,
$\vec{b}'_{m+1}$ has at least one more standard component that
$\vec{b}'_m$.  Since clearly $\nu = \sum_{j=0}^m\epsilon_j \vec{b}_j +
\epsilon_{m+1} \vec{b}_{m+1}'$, this completes the inductive step.
The lemma follows immediately.
\eprf

Returning to the proof of Theorem~\ref{lpsnps}, 
I next prove by induction on $m$ that for all $m \le k'$ (where $k' \le
k$ is as in Lemma~\ref{newlem1}), there exist standard probability measures 
$\mu_0, \ldots, \mu_m$,  (standard) vectors $\vec{b}_{m+1},
\ldots, \vec{b}_{k'} \in \IR^k$, and $\epsilon_1, \ldots,
\epsilon_{k'}$ such that $\nu = \sum_{j=0}^m \epsilon_j \mu_j +
\sum_{j = m+1}^{k'} \epsilon_j \vec{b}_j$.

The base case is immediate from Lemma~\ref{newlem1}: taking $\vec{b}_j$,
$j = 1, \ldots, k'$ as in Lemma~\ref{newlem1}, 
$\vec{b}_0$ is in fact a probability measure since $\vec{b}_0 = \stand{\nu}$.
Suppose that the result holds for $m$. Consider $\vec{b}_{m+1}$.
If $b_{(m+1)i} < 0$ for some $j$ then, since $\nu(U_i) \ge 0$,  
there must exist $j' \in \{1, \ldots, m\}$ such that $\mu_{j'}(U_i) >
0$.  Thus, there exists some $N > 0$ such that $N(\mu_{j'}(U_i)) +
b_{(m+1)i} > 0$.
Since there are only finitely many basic elements
and every element in the vector $\mu_j$ is nonnegative, for $j = 0,
\ldots, m$, there must exist some
$N'$ such that
$\vec{b}'_{m+1} = N'( \mu_0 + \cdots
+  \mu_m) + \vec{b}_{m+1} \ge 0$.  Let $c = \sum_{i = 1}^k b_{(m+1)i}'$, and
let $\mu_{m+1} = \vec{b}'_{m+1}/c$.  Clearly, 
$\nu = (\epsilon_0 -N' \epsilon_{m+1}) \mu_0 + \cdots (\epsilon_m - N'
\epsilon_{m+1}) \mu_m + c \epsilon_{m+1} \mu_{m+1} + \sum_{j=m+2}^{k'}
\vec{b}_j$.  This completes the proof of the inductive step.

The theorem now immediately follows. \eprf

\bigskip

\opro{infiniteeq} Every LPS over $(W,\F)$ is
equivalent to an LPS over $(W,\F)$ of length at most $|\F|$. \eopro

\medskip

\prf The argument is essentially the same as that for
Proposition~\ref{finiteeq}, using the observation that 
a probability measure over $(W,\F)$ can be identified with an element of
$\IR^{|\F|}$; the vector defining the probabilities of the elements in
$\F$.  I leave details to the reader. \eprf

\pro\label{counter} For the NPS $(W,\F,\nu)$ constructed in
Example~\ref{counter4}, 
there is no LPS $\vecmu$ over $(W,\F)$ such that $\nu \aeq
\vecmu$. \epro

\prf I start with a straightforward lemma.

\lem\label{distinct} Given an LPS $\vecmu$, there is an LPS $\vecmu'$
such that $\vecmu 
\aeq \vecmu'$ and all the probability measures in $\vecmu'$ are
distinct.
\elem

\prf Define $\vecmu'$ to be the subsequence consisting of all the
distinct probability measures in $\vecmu$.  That is, suppose that $\vecmu =
(\mu_0, \mu_1, \ldots )$.  Then $\vecmu' = (\mu_{k_0}, \mu_{k_1},
\ldots )$, where $k_0 = 0$, and, if $k_\alpha$ has been defined for all
$\alpha < \beta$ and
there exists an index $\gamma$ such that $\mu_{k_\alpha} \ne \mu_\gamma$ for
all $\alpha \le \beta$, then $k_\beta$ is the least index $\delta$ such that 
$\mu_{k_\alpha} \ne \mu_\delta$ for all $\alpha < \beta$.  If there is no
index $\gamma$ such that $\mu_\gamma \notin \{\mu_{k_\alpha}: \alpha <
\beta\}$, then $\vecmu' = (\mu_{k_\alpha}: \alpha < \beta)$.  I leave
it to the reader to check that $\vecmu \aeq \vecmu'$. \eprf

Returning to the proof of Proposition~\ref{counter}, suppose by way of
contradiction that $\nu \aeq \vecmu$.  Without loss of generality, by
Lemma~\ref{distinct}, assume that all the probability measures
in $\vecmu$ are distinct.
Clearly
$E_\nu(\chi_W) < E_\nu(\alpha \chi_{\{w_1\}})$ if $\alpha \ge 2$ and
$E_\nu(\chi_W) >
E_\nu(\alpha \chi_{\{w_1\}})$ if $\alpha < 2$.  Since $\nu \aeq \vecmu$,
it must be 
the case that $E_{\vecmu}(\chi_W) < E_{\vecmu}(\alpha \chi_{\{w_1\}})$ if
$\alpha \ge 2$ 
and $E_{\vecmu}(\chi_W) >  
E_{\vecmu}(\alpha \chi_{\{w_1\}})$ if $\alpha < 2$.  Since $E_{\vecmu}(\chi_W)
= (1, 1, \ldots)$, it follows that if $\vecmu = (\mu_0, \mu_1, \ldots)$,
it must 
be the case that $\mu_0(w_1) = 1/2$ and 
\begin{equation}\label{eq:mu1}
\mu_1(w_1) \ge 1/2.
\end{equation}
Similar
arguments (comparing $\chi_W$ to $\chi_{\{w_{j}\}}$) can be used to show that
$\mu_0(w_j) = 1/2^j$ and $\mu_1(w_{2j-1}) \ge 1/2^j$ for $j = 1, 2,
\ldots$. 
Next, observe that $E_{\nu}(\chi_{\{w_1\}} - 2^{2k-1}\chi_{\{w_{2k}\}}) =
(2^{k} + 1)\epsilon$.  Thus, $$E_{\nu}(\chi_{\{w_1\}} -
2^{2k-1}\chi_{\{w_{2k}\}}) = E_{\nu}((2^{k}+1)(\chi_{\{w_1\}} - (\chi_W/2))).$$
It follows that the same relationship must hold if $\nu$ is replaced by
$\vecmu$.  That is, 
$$\mu_1(w_1) - 2^{2k-1}\mu_1(w_{2k}) =
(2^{k}+1)(\mu_1(w_1) - (1/2)).$$
Rearranging terms, this gives
$$2^{k}\mu_1(w_1) + 2^{2k-1}\mu_1(w_{2k}) = 2^{k-1} + 1/2,$$
or
\begin{equation}\label{eq1.5}
\mu_1(w_1) + 2^{k-1} \mu_1(w_{2k}) = 1/2 + 1/2^{k+1}.
\end{equation}
Thus, $\mu_1(w_1) \le 1/2 + 1/2^{k+1}$ for all $k \ge 1$.
Putting this together with (\ref{eq:mu1}), it
follows that $\mu_1(w_1) = 1/2$.  Plugging this into (\ref{eq1.5}) gives
$\mu_1(w_{2k}) = 1/2^{2k}$.  It now follows that $\mu_1 =
\mu_0$, contradicting the choice of $\vecmu$.  \eprf

\bigskip

\othm{FNP} $\FNP$ is a bijection from $\NPS(W,\F)/\!\simeq$ to
$\Popper(W,\F)$ and from $\NPS^c(W,\F)/\!\simeq$ to $\Popper^c(W,\F)$.
\eothm

\medskip

\prf 
As I said in the main text, the proof that $\FNP$ is an injection is
straightforward, and to prove that it is a surjection in the countably
additive case, it suffices to show that $\FNP(W,\F,\nu) =
(W,\F,\F',\mu)$, where $\nu \aeq \vecmu'$ and $\vecmu'$ is the
countably additive SLPS such that $\FCP((W,\F,\vec{\mu}'))
= (W, \F,\F', \mu)$.   I now do this.

Suppose that $\FNP(W,\F,\nu) = (W,\F,\F_1',\mu_1)$. 
First I show that $\nu(U) = 0$ iff $\vecmu'(U) = \vec{0}$.  
Let $X = \chi_U$ and $Y = \chi_{\emptyset}$.  Note that $\nu(U) = 0$ iff
$E_\nu(X) = E_\nu(Y)$ iff $E_{\vecmu'}(X) = E_{\vecmu'}(Y)$ iff
$\vecmu'(U) = \vec{0}$.  Thus, $\F_1' = \{U: \nu(U) \ne 0\} = 
\{U: \vecmu'(U) \ne \vec{0}\} = \F'$. 

Now suppose by way of contradiction that $\mu \ne \mu_1$.  Thus, there
must exist some $V \in \F$, $U \in \F'$ such that $\mu(V \mid U) \ne
\mu_1(V \mid U)$.  Let $\beta$ be the smallest ordinal such that
$\mu_\beta'(U) \ne 0$.  It follows that $\mu'_\beta(V \mid U) \ne \stand{\nu(V
\mid U)}$.  We can assume without loss of generality that $\mu'_\beta(V
\mid U) > \stand{\nu(V \mid U)}$.  Choose a real number $r$ such that  
$\mu'_\beta(V \mid U) > r >  \stand{\nu(V \mid U)}$.  Then
$E_{\vecmu'}(\chi_{V \inter U}) > E_{\vecmu'}(r \chi_U)$ but
$E_{\nu}(\chi_{V \inter U}) < E_{\nu}(r \chi_U)$.  This contradicts the
assumption that $\vecmu' \aeq \nu$.  It follows that $\FNP(W,\F,\nu) =
(W,\F,\F',\mu)$, as desired.

It remains to show that if $(W,\F,\F',\mu) \in \Popper(W,\F) -
\Popper^c(W,\F)$, then there is some $(W,\F,\nu) \in \NPS(W,\F)$ such that
$\FNP(W,\F\nu) = (W,\F,\F',\mu)$.  My proof in this case follows closely
the lines of 
an analogous result proved by 
McGee \citeyear{McGee94}.  I provide the details here mainly for
completeness.  

The proof relies on the following ultrafilter construction of
non-Archimedean fields.  Given a set $S$, a {\em filter\/} $\G$ on $S$ is a
nonempty set of subsets of $\F$ that is closed under supersets (so that
if $U \in \G$ and $U \subseteq U'$, then $U' \in \G$), is closed under
finite intersections (so that if $U_1, U_2 \in \G$, then $U_1
\inter U_2 \in \G$), and does not contain $\emptyset$.  An {\em
ultrafilter\/} is a maximal filter, that is, a filter that is not a
strict subset of any other filter.  It is not hard to show that if $\U$
is an ultrafilter on $S$, then for all $U \subseteq S$, either $U \in
\U$ or $\overline{U} \in \U$ \cite{BellSlomson}.

Suppose $F$ is either $\IR$ or a 
non-Archimedean field, $J$ is an arbitrary set, and $\U$ is an
ultrafilter on $J$.  Define an equivalence relation $\sim_{\U}$ on
$F^J$ by taking $(a_j: j \in J) \sim_{\U} (b_j: j \in J)$ if $\{j: a_j =
b_j\} \in \U$.  Similarly, define a total order $\preceq_\U$ by taking
$(a_j: j \in J) \preceq_{\U} (b_j: j \in J)$ if $\{j: a_j \le b_j\} \in
\U$.  (The fact that $\le_{\U}$ is total uses the fact that for all $U
\subseteq 
J$, either $U \in \U$ or $\overline{U} \in \U$.  Note that the pointwise
ordering on $F^J$ is not total.)  Let $F^J/\nsim_{\U}$ consist of these
equivalence classes.  Note that $F$ can be viewed as a subset of
$F^J/\nsim_{\U}$ by identifying $a \in F$ with the sequence of all $a$'s.

Define addition and multiplication on $F^J$ pointwise,
so that, for example, $(a_j: j \in J) + (b_j: j \in J) = (a_j + b_j: j
\in J)$.  It is easy to check that if $(a_j: j \in J) \sim_{\U} (a_j': j
\in J)$, then $(a_j: j \in J) + (b_j: j \in J) \sim_{\U} (a_j': j \in J) +
(b_j: j \in J)$, and similarly for multiplication.  Thus, the
definitions of $+$ and $\times$ can be extended in the obvious way to
$F^J/\nsim_{\U}$.  With these definitions, it is easy to check that
$F^J/\nsim_{\U}$ is a field that contains $F$.

Now given a Popper space $(W,\F,\F',\mu)$ and a finite subset $\A = \{U_1,
\ldots, U_k\} \subseteq \F$, let $\F_{\A}$ be the (finite) algebra
generated by $\A$ (that is, the smallest set containing $\{U_1, \ldots,
U_k, W\}$ that is closed under unions and complement).  Let
$\F'_{\A} = \F_{\A} \inter \F'$.  It follows from Theorem~\ref{FCPfin} that
there is a finite SLPS $\vecmu_\A$ over $(W,\F_{\A})$ that is mapped to
$(W,\F_{\A},\F'_{\A'}, \mu_{\A})$ by $\FCP$.  (Although
Theorem~\ref{FCPfin} is stated for finite state spaces $W$, the proof
relies on only the fact that the algebra is finite, so it applies without
change here.)  It now follows from
Theorem~\ref{lpsnps} that, for each $\A$, there is a nonstandard
probability space $(W,\F_{\A},\nu_\A)$ with range $\IR(\epsilon)$ that is
equivalent to $\vecmu_{\A}$.  By Proposition~\ref{FCPaeq}, it follows
that for $U \in \F'_{\A}$ iff $\nu_{\A}(U) = 0$.
Moreover, $\stand{\nu_{\A}(V \mid U)} = \mu_{\A}(V \mid U)$ for $U \in
\F'_{\A}$ and $V \in \F_{\A}$.

Let $J$ consist of all finite subsets of $\F$.  For a subset $\A$ of
$\F$, let $G_{\A}$ be the subset of $2^J$ consisting of all sets in $J$
containing $\A$.  Let $\G = \{G \subseteq J: G \supseteq G_{\A} \mbox{ for
some } \A \subseteq \F\}$.  It is easy to check that $\G$ is a filter on
$J$.  It is a standard result that every filter can be extended to an
ultrafilter \cite{BellSlomson}.  Let $\U$ be an ultrafilter containing
$\G$.  By the construction above, $\R(\epsilon)/\nsim_{\U}$ is a
non-Archimedean field.

Define $\nu$ on $(W,\F)$ by taking
$\nu(U) = (\nu_{\A}(U): \A \in J)$, where $\nu_\A(U)$ is taken to be 0
if $U \notin \F_{\A}$.  To see that $\nu$ is indeed a nonstandard
probability measure with the required properties, note that clearly
$\nu(W) = 1$ (where 1 is identified with the sequence of all 1's).
Moreover, to see that $\nu(U) + \nu(V) = \nu(U \union V)$, let
$\A_{U,V}$ be the smallest subalgebra containing $U$ and $V$.
Note that if $\A \supset \A_{U,V}$, then 
$\nu_{\A}(U) + \nu_{\A}(V) = \nu_{\A'}(U \union V)$.  Since the set of
algebras containing $\A_{U,V}$ is an element of the ultrafilter, the
result follows.  Similar arguments show that $\nu(U) = 0$ iff $U \in
\F'$ and that $\stand{\nu(V \mid U)} = \mu(V \mid U)$ if $U \in \F'$ and $V \in
\F$. Clearly $\FNP(\nu) = \mu$.   \eprf

\bigskip

\opro{simeqvsaeq} If $\nu_1 \aeq \nu_2$ then $\nu_1 \simeq \nu_2$. 
\eopro

\medskip

\prf Suppose that $\nu_1 \aeq \nu_2$.  To show that $\nu_1 \simeq
\nu_2$, first suppose that $\nu_1(U) \ne 0$ for some $U \subseteq W$.  Then
$E_{\nu_1}(\chi_\emptyset) < E_{\nu_1}(\chi_U)$.  Since $\nu_1 \aeq
\nu_2$, it must be the case that $E_{\nu_2}(\chi_\emptyset) <
E_{\nu_2}(\chi_U)$.  Thus, $\nu_2(U) \ne 0$.  A symmetric argument shows
that if $\nu_2(U) \ne 0$ then $\nu_1(U) \ne 0$.  Next, suppose that
$\nu_1(U) \ne 0$ and $\nu_1(V \mid U) = \alpha$.  Thus,
$E_{\nu_1}(\alpha \chi_U) = E_{\nu_1}(\chi_{U \inter V})$.  Since 
$\nu_1 \aeq \nu_2$, it follows that
$E_{\nu_2}(\alpha \chi_U) = E_{\nu_2}(\chi_{U \inter V})$, and so 
$\nu_2(V \mid U) = \alpha$.  Thus, $\stand{\nu_1(V \mid U)} =
\stand{\nu_2(V \mid  U)}$.
Hence, $\nu_1 \simeq \nu_2$, as desired. \eprf
\commentout{
\bigskip

\opro{indaeq}
$U$ is approximately conditionally independent of $V$
given $V'$ with respect to $\nu$ iff there exists a measure $\nu'$ such
that $\nu \aeq \nu'$ and $U$ is conditionally independent of $V$ given
$V'$ with respect to $\nu'$.
\eopro

\medskip

\prf Suppose that $U$ is approximately conditionally independent of $V$
given $V'$ with respect to $\nu$.   If $\nu(U \inter V') = 0$, then $U$
is conditionally independent of $V$ given $V'$ with respect to $\nu$.
If $\nu(U \inter V') \ne 0$, $\stand{\nu(V \mid U \inter V')}
= \stand{\nu(V \mid V')}$.  
}

\othm{BBDstrongindependence}  There exists an NPS $\nu$ whose
range is an
elementary extension of the reals such that $\vecmu \aeq \nu$ and $X_1,
\ldots, X_n$ are  independent with respect to $\nu$ iff there
exists a sequence $\vec{r}^j$, $j = 1, 2, \ldots$ of vectors in $(0,1)^k$ 
such that $\vec{r}^j \rightarrow (0,\ldots, 0)$ as $j\rightarrow\infty$, 
and $X_1, \ldots, X_n$ are independent with respect to $\vecmu \, \Box
\, \vec{r}^j$ for $j = 1, 2, 3, \ldots$.   
\eothm

\prf Suppose that  there exists an NPS 
$\nu$ whose range is an elementary extension of the reals, $\vecmu
\aeq \nu$,  and $X_1, \ldots, X_n$ are 
independent with respect to $\nu$.  Using arguments similar in spirit to
those the 
arguments of BBD \citeyear[Proposition 2]{BBD2}, it follows that there exist
positive infinitesimals $\epsilon_1, \ldots, \epsilon_k$ such that
$\vecmu \, \Box \, (\epsilon_1, \ldots, \epsilon_k) = \nu$.  It is not
hard to show that there exist a finite set of real-valued polynomials
$p_1,\ldots, p_N$ such that $p_j(\epsilon_1, \ldots, \epsilon_k) = 0$
for $j = 1, \ldots, N$ and if $\vec{r}$ is a vector of positive reals
such that $p_j(\vec{r}) = 0$ for $j = 1, \ldots, N$, then $X_1, \ldots,
X_n$ are independent with respect to $\vecmu \, \Box \, \vec{r}$.
Thus, for all natural numbers $m \ge 1$, the range of 
$\nu$ satisfies the first-order property $$\exists x_1 \ldots \exists x_k 
(p_1(x_1, \ldots, x_k) = 0 \land \ldots \land p_N(x_1, \ldots, x_k) = 0
\land 0 < x_1 < 1/m \land \ldots \land 0 < x_k < 1/m).$$
Since the range of $\nu$ is an elementary extension of the reals, this
first-order 
property holds of the reals as well.
Thus, there exists a sequence
$\vec{r}^j$ of vectors of positive reals converging to $\vec{0}$ such that
$p_j(\vec{r}^j) = 0$ for $j = 1, \ldots, N$.  

The converse follows by a straightforward application
of compactness in first-order logic \cite{Enderton}.  
Suppose that there exists a sequence 
$\vec{r}^j$, $j = 1, 2, \ldots$ of vectors in $(0,1)^k$ 
such that $\vec{r}^j \rightarrow (0,\ldots, 0)$ 
as $j\rightarrow\infty$, and $X_1, \ldots, X_n$ are
independent with respect to $\vecmu \, \Box \, \vec{r}^j$ for $j = 1, 2, 3,
\ldots$.   We now apply the compactness theorem.  
As I mentioned in the proof of Proposition~\ref{infiniteeq}, the
compactness theorem says that, 
given a collection for formulas, if each finite subset has a model, then
so does the whole set.  
Consider a language with the function symbols $+$ and $\times$, 
the binary relation $\le$, a constant
symbol $\mathbf{r}$ for each 
real number $r$, a unary predicate $N$ (representing the natural numbers),
and constant symbols $p_{U}$ for each set $U \in 
\F$.  Intuitively, $p_U$ represents $\nu(U)$.
Consider the following (uncountable) collection of formulas:
\begin{itemize}
\item[(a)]  All first-order formulas in this language true of the reals.
(This includes, for example, a formula such as $\forall x\forall y(x+ y
=  y+x)$, which says that addition is commutative, as well as formulas
such as $\mathbf{2} + \mathbf{3} = \mathbf{5}$ and 
$\mathbf{\sqrt{2}} \times \mathbf{\sqrt{3}} = \mathbf{\sqrt{6}}$.)
\item[(b)] Formulas $p_U > 0$ for $U \in \F'$ and $p_U = 0$ for $U \in \F -
\F'$.
\item[(c)] Formulas $p_U + p_V = p_{U \union V}$ if $U \inter V = \emptyset$.
\item[(d)] The formula $p_W = 1$.
\item[(e)] Formulas of the form $p_{X_1 = x_1} \times \cdots \times
p_{X_n = x_n} = 
p_{X_1 = x_1 \inter \ldots \inter X_n = x_n}$, for all values $x_i \in
\V(X_i)$, $i = 1, \ldots, n$; these formulas say that $X_1,
\ldots, X_n$ are independent with respect to $\nu$.
\item[(f)] For every pair of $Y$, $Y'$ of random variables such that
$E_{\vecmu}(Y) \ge E_{\vecmu}(Y')$, a formula that says
$E_{\nu}(Y) \ge E_{\nu}(Y')$, where $E_{\nu}(Y)$ and $E_{\nu}(Y')$ are
expressed using the constant symbols $p_U$ (where the events $U$ are
those of the form $Y=y$ and $Y'=y'$).  
Note that this formula is finite, since $X$ and $Y$ are assumed to have
finite range.  The formula would not be expressible in first-order logic
if $X$ or $Y$ had infinite range.
\end{itemize}

It is not hard to show that every finite subset of these formulas is
satisfiable.  Indeed, given a finite subset of formulas, there must
exist some $m$ such that taking $p_U = \vecmu \, \Box \, \vec{r}^m(U)$
will work (and interpreting $\mathbf{r}$ as the real number $r$, of
course).  The only nonobvious part is showing that we can deal with the
formulas in part (f); that we can do so follows from the proof of
Proposition 1 in  \cite{BBD2}, which shows that 
$E_{\vecmu}(Y') > E_{\vecmu}(Y)$ iff there exists some $M$ such that $E_{\vecmu \, \Box \,
\vec{r}^m}(Y') >  
E_{\vecmu \, \Box \, \vec{r}^m}(Y)$ for all $m$, then 
$E_{\vecmu}(Y') > E_{\vecmu}(Y)$.  

Since every finite set of formulas is satisfiable, 
by compactness, the infinite set is satisfiable.  Let $\nu(U)$
be the interpretation of $p_U$ in a model satisfying these formulas.
Then it is easy to check that $\nu$ is an elementary extension of the
reals, $\nu \aeq \vecmu$, and  
that $X_1, \ldots, X_n$ are independent with respect to $\nu$.
\eprf

\othm{KRindependence}  
$X_1, \ldots, X_n$ are strongly independent with respect to the Popper
space $(W,\F,\F',\mu)$ iff there 
exists an NPS $(W,\F,\nu)$ such that 
$\FNP(W,\F,\nu) = (W,\F,\F',\mu)$ and $X_1, \ldots,
X_n$ are independent with respect to $(W,\F,\nu)$.  
\eothm

\prf  It easily follows from Kohlberg and Reny's \citeyear[Theorem
2.10]{KR97} characterization of strong independence that if 
$X_1, \ldots, X_n$ are independent with respect to the NPS
$(W,\F,\nu)$ then $X_1, \ldots, X_n$ are strongly independent with respect to
$\FNP(W,\F,\nu)$. 
\commentout{
The converse follows by a straightforward application
of compactness in first-order logic \cite{Enderton}.  

Suppose that $(W,\F,\F',\mu)$ is a Popper space and 
$\mu_j \rightarrow \mu$ are as required for $X_1, \ldots, X_n$ to be
strongly independent with respect to $\mu$.
As I mentioned in the proof of Proposition~\ref{infiniteeq}, the
compactness theorem says that, 
given a collection for formulas, if each finite subset has a model, then
so does the whole set.  
Consider a language with the function symbols $+$ and $\times$, 
the binary relation $\le$, a constant
symbol $\mathbf{r}$ for each 
real number $r$, and constant symbols $p_{U}$ for each set $U \in 
\F$.  Intuitively, $p_U$ represents $\nu(U)$.
Consider the following (uncountable) collection of formulas:
\begin{itemize}
\item All formulas true in fields (for example, $\forall x, y (x+ y =
y+x)$, which says that addition is commutative).
\item All true statements of the form $\mathbf{r_1} + \mathbf{r_2} =
\mathbf{r_3}$ and $\mathbf{r_1} \times \mathbf{r_2} = \mathbf{r_3}$
involving real constants $\mathbf{r_1}$, $\mathbf{r_2}$, $\mathbf{r_3}$
(for example $\mathbf{2} + \mathbf{3} = \mathbf{5}$ and 
$\mathbf{\sqrt{2}} \times \mathbf{\sqrt{3}} = \mathbf{\sqrt{6}}$).
\item Formulas $p_U > 0$ for $U \in \F'$ and $p_U = 0$ for $U \in \F -
\F'$.
\item Formulas $p_U + p_V = p_{U \union V}$ if $U \inter V = \emptyset$.
\item The formula $p_W = 1$.
\item Formulas of the form $p_{X_1 = x_1} \times \cdots \times p_{X_n = x_n} =
p_{X_1 = x_1 \inter \ldots \inter X_n = x_n}$, for all values $x_i \in
\V(X_i)$, $i = 1, \ldots, n$; these formulas say that $X_1,
\ldots, X_n$ are independent with respect to $\nu$.
\item Formulas of the form $(\mathbf{r - \frac{1}{n}})p_V  \le p_{U \inter V}
\le (\mathbf{r + \frac{1}{n}})p_V$ for all $U$, $V$, $\mathbf{r}$, and
$\mathbf{n} > 0$ such that $\mu(U \mid V) = r$.
\end{itemize}

It is easy to see that every finite subset of these formulas is
satisfiable.  Indeed, given a finite subset of formulas, there must
exist some $m$ such that taking $p_U = \mu_m(U)$ satisfies all the
formulas (and interpreting $\mathbf{r}$ as the real number $r$, of
course).  By compactness, the infinite set is satisfiable.  Let $\nu(U)$
be the interpretation of $p_U$ in a model satisfying these formulas.
Then it is easy to check that $\FLN(W,\F,\nu) = (W,\F,\F',\mu)$, 
and that $X_1, \ldots, X_n$ are independent with respect to $\nu$.
}

The converse follows using compactness, much as in the proof of
Theorem~\ref{BBDstrongindependence}.
Suppose that $(W,\F,\F',\mu)$ is a Popper space and 
$\mu_j \rightarrow \mu$ are as required for $X_1, \ldots, X_n$ to be
strongly independent with respect to $\mu$.
Consider the same language as in the proof of
Theorem~\ref{BBDstrongindependence}, and essentially the same
collection of formulas, except that the formulas of part (f) are
replaced by 
\begin{itemize}
\item[(f$'$)] Formulas of the form $(\mathbf{r - \frac{1}{n}})p_V  \le p_{U \inter V}
\le (\mathbf{r + \frac{1}{n}})p_V$ for all $U$, $V$, $\mathbf{r}$, and
$\mathbf{n} > 0$ such that $\mu(U \mid V) = r$.
\end{itemize}

Again, it is easy to see that every finite subset of these formulas is
satisfiable.  Indeed, given a finite subset of formulas, there must
exist some $m$ such that taking $p_U = \mu_m(U)$ satisfies all the
formulas (and interpreting $\mathbf{r}$ as the real number $r$, of
course).  By compactness, the infinite set is satisfiable.  Let $\nu(U)$
be the interpretation of $p_U$ in a model satisfying these formulas.
Then it is easy to check that $\FLN(W,\F,\nu) = (W,\F,\F',\mu)$, 
and that $X_1, \ldots, X_n$ are independent with respect to $\nu$.
\eprf

\bibliographystyle{chicago}
\bibliography{z,joe}
\end{document}